\documentclass[%
superscriptaddress,
amsmath,amssymb,
aps,
prl,
floatfix,
longbibliography
]{revtex4-1}
\usepackage[utf8]{inputenc}
\usepackage{graphicx}
\usepackage{dcolumn}
\usepackage{bm}
\usepackage{hyperref}
\usepackage{amsmath,amsfonts,amssymb} 
\usepackage{pdfpages}
\usepackage{float}
\usepackage{braket}
\usepackage{tikz}
\usepackage{CJK}
\usepackage{siunitx}
\usepackage{xcolor}
\usepackage{tabularx}
\makeatletter
\AtBeginDocument{\let\LS@rot\@undefined}
\makeatother

\newcommand{\VT}{\ensuremath V_\mathrm{t}}  \newcommand{\VB}{\ensuremath V_\mathrm{b}}
   
\newcommand{\Vtop}{\ensuremath V_\mathrm{tBG}}  \newcommand{\Vbottom}{\ensuremath V_\mathrm{bBG}}
\newcommand{\ntot}{\ensuremath n_\mathrm{tot}}   \newcommand{\nout}{\ensuremath n_\mathrm{out}}
\newcommand{\nt}{\ensuremath n_\mathrm{t}}  \newcommand{\nb}{\ensuremath n_\mathrm{b}}
   
\newcommand{\Ct}{\ensuremath C_\mathrm{t}}  \newcommand{\Cb}{\ensuremath C_\mathrm{b}}
\newcommand{\Cqt}{\ensuremath C_\mathrm{qt}}  \newcommand{\Cqb}{\ensuremath C_\mathrm{qb}}
\newcommand{\Cqte}{\ensuremath C_\mathrm{qte}}  \newcommand{\Cqbe}{\ensuremath C_\mathrm{qbe}}
\newcommand{\Cqth}{\ensuremath C_\mathrm{qth}}  \newcommand{\Cqbh}{\ensuremath C_\mathrm{qbh}}
\newcommand{\Cqe}{\ensuremath C_\mathrm{qe}}  \newcommand{\Cqh}{\ensuremath C_\mathrm{qh}}
\newcommand{\Cq}{\ensuremath C_\mathrm{q}}
\newcommand{\Cm}{\ensuremath C_\mathrm{m}}

\newcommand{\epsilonhBN}{\ensuremath \epsilon_\mathrm{hBN}}  \newcommand{\epshBN}{\ensuremath \epsilon_\mathrm{hBN} \epsilon_0}
\newcommand{\eps}{\ensuremath \epsilon_0}

\newcommand{\DT}{\ensuremath D_\mathrm{t}}  \newcommand{\DB}{\ensuremath D_\mathrm{b}}
  
\newcommand{\cfieldT}{\ensuremath \mathcal{E}_\mathrm{t}}  \newcommand{\cfieldB}{\ensuremath \mathcal{E}_\mathrm{b}}
\newcommand{\dT}{\ensuremath d_\mathrm{t}}  \newcommand{\dB}{\ensuremath d_\mathrm{b}}
\newcommand{\dBG}{\ensuremath d_\mathrm{g}}
\newcommand{\EFbottom}{\ensuremath E_\mathrm{F,b}}  
\newcommand{\Dose}{\ensuremath\mathcal{D}_\mathrm{e}}  \newcommand{\Dosh}{\ensuremath\mathcal{D}_\mathrm{h}}  
\newcommand{\me}{\ensuremath m^*_\mathrm{e}}
\newcommand{\mh}{\ensuremath m^*_\mathrm{h}}
\newcommand{\melectron}{\ensuremath m_\mathrm{e}}

\begin{document}
	\title{Gap Opening in Twisted Double Bilayer Graphene by Crystal fields}

	\author{Peter Rickhaus}
	\author{Giulia Zheng}
	\affiliation{Solid State Physics Laboratory, ETH Zürich,~CH-8093~Zürich, Switzerland}
	\author{Jose L. Lado}
	\affiliation{Department of Applied Physics, Aalto University, Espoo, Finland}
	\affiliation{Institute for Theoretical Physics, ETH Zurich, 8093 Zurich, Switzerland}
	\author{Yongjin Lee}
	\author{Annika Kurzmann}
	\author{Marius Eich}
	\author{Riccardo Pisoni}
	\author{Chuyao Tong}
	\author{Rebekka Garreis}
	\author{Carolin Gold}
	\author{Michele Masseroni}
	\affiliation{Solid State Physics Laboratory, ETH Zürich,~CH-8093~Zürich, Switzerland}
	\author{Takashi Taniguchi}
	\author{Kenji Wantanabe}
	\affiliation{National Institute for Material Science, 1-1 Namiki, Tsukuba 305-0044, Japan}
	\author{Thomas Ihn}
	\author{Klaus Ensslin}
	\affiliation{%
		Solid State Physics Laboratory, ETH Zürich,~CH-8093~Zürich, Switzerland}
	
	\date{\today}
	
	\begin{abstract}
		Crystal fields occur due to a potential difference between chemically different atomic species. 
		In Van-der-Waals heterostructures such fields are naturally present perpendicular to the planes. It has been realized recently that twisted graphene multilayers provide powerful playgrounds to engineer electronic properties by the number of layers, the twist angle, applied electric biases, electronic interactions and elastic relaxations, but crystal fields have not received the attention they deserve. Here we show that the bandstructure of large-angle twisted double bilayer graphene is strongly modified by crystal fields. In particular, we experimentally demonstrate that twisted double bilayer graphene, encapsulated between hBN layers, exhibits an intrinsic bandgap. By the application of an external field, the gaps in the individual bilayers can be closed, allowing to determine the crystal fields. We find that crystal fields point from the outer to the inner layers with strengths in the bottom/top bilayer $\cfieldB=\SI{0.13}{V/nm}\approx-\cfieldT=\SI{0.12}{V/nm}$.  We show both by means of first principles calculations and low energy models that crystal fields open a band gap in the groundstate. Our results put forward a physical scenario in which a crystal field effect in carbon substantially impacts the low energy properties of twisted double bilayer graphene, suggesting that such contributions must be taken into account in other regimes to faithfully predict the electronic properties of twisted graphene multilayers.
	\end{abstract}
	\maketitle


	\section{Introduction}
	
	\begin{figure}
		\centering
		\includegraphics[width=1\textwidth]{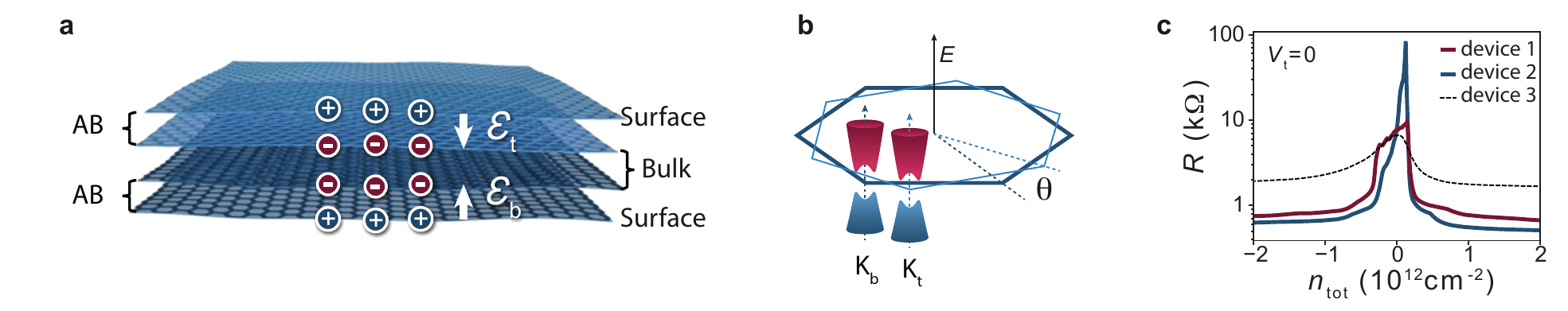}
		\caption{a) Two twisted, AB-stacked bilayer graphene (BG) sheets. The electrostatic potential of the outer layers is different from the potential in the inner layers. This leads to crystal fields $\cfieldT=-\cfieldB$ pointing in opposite direction in the top and bottom BG. In the experiment, the two bilayer systems are encapsulated in hBN which reduces the strength of the crystal fields compared to vacuum.
			b) The TDBG band structure consists of Brillouin-zones of the top and bottom BG rotated with respect to each other. For large twist angles $\theta$, the bands of the top and bottom layer intersect at energies large compared to the Fermi energies of the individual layers. Therefore, at typical Fermi energies, the individual BG band structures remain intact. The crystal fields open a single-particle gap in both layers.
			c) In such a structure we observe a gap at zero density and zero external field in a resistance versus density trace $R(\ntot)$. We show traces for two devices, device 1 is further discussed in the main text and further measurements of device 2 and device 3 are shown in the Supplemental Material.  
		}\label{fig:1}
	\end{figure}
	
	In a three-dimensional crystal, the electrostatic potential at the surface or at the interface to other materials differs from the potential in the bulk leading to crystal fields which may influence the bandstructure near the interface.
	In two-dimensional Van-der-Waals heterostructures the bulk consists of one or only few atomic layers. Therefore, crystal fields can significantly alter the band structure of the entire crystal. The crystal field contribution has been theoretically predicted for graphitic systems \cite{Gruneis2008} and twisted graphene multilayers \cite{2019arXiv190600623H} and can yield a substantial modification of the band structure of multilayer heterostructures with more than two layers, but so far it has not been possible to probe such modification in a systematic and quantitative way.
	This becomes possible by using twisted double-bilayer graphene (TDBG) with a large twist angle between the Bernal stacked bilayer graphene (BG) layers (Fig. \ref{fig:1}a).
	On the one hand the two BG layers are electronically decoupled by a large momentum mismatch \cite{Rozhkov2016} of the K-points in the bottom ($K_\mathrm{b}$) and the top layer ($K_\mathrm{t}$) as shown in Fig. \ref{fig:1}b. Such decoupling has been experimentally observed for large-angle twisted graphene in the quantum Hall regime \cite{Lucian2011,Sanchez2012} as well as at zero magnetic field using Fabry-Pérot resonators \cite{Rickhaus2019}. Therefore, at low enough energies, the band structure of TDBG can be described by the well-established theories for bilayer graphene \cite{McCann2006,Castro2007}.
	On the other hand, each of the BG layers is sensitive to an out-of-plane electric field $D$. At finite $D$, the BG band structure becomes gapped \cite{Oostinga2007,Zhang2009}. In experiments, this single-particle gap is induced and tuned using dual-gated geometries. This has been exploited to define electrostatic nanostructures such as quantum point contacts \cite{Goossens2012,Overweg2017a} and quantum dots \cite{Allen2012,Eich2018}. 
	With its four layers, TDBG is thin enough that crystal fields matter. With its bilayer graphene character it is sensitive to crystal fields appearing between the surface (outer two layers) and the bulk (inner two layers) (see Fig.\ref{fig:1}a).
	
	Here we demonstrate experimentally and theoretically that the ground state of TDBG possesses a single-particle gap , induced entirely by crystal fields. We fabricate TDBG devices with large twist angles (device 1: $\theta=\SI{15}{\degree}$, device 2:  $\theta=\SI{30}{\degree}$, device 3: $\theta=\SI{10}{\degree}$) and measure their low temperature ($T=\SI{1.5}{K}$) transport properties. The devices are tuned with a global graphite back gate ($\VB$) and a narrow metallic top gate ($\VT$) with the exception of device 3 which has a global top gate. 
	By changing $\VT$ and $\VB$ we can tune the densities in the top ($\nt$) and bottom BG ($\nb$) individually. The resulting conductance map shows distinct zero density lines for both BG layers. These lines are in agreement with an electrostatic model that takes into account quantum capacitances. We extract a geometric capacitance between the BG layers ($\Cm=\SI[separate-uncertainty = true]{3.5\pm 1}{\micro F/cm^2}$) which is half the observed capacitance between two twisted single-layer graphene sheets \cite{Rickhaus2019}. At zero applied voltage ($\VT=\VB=0$) we observe a strong resistance peak (Fig.\ref{fig:1}c) caused by an energy gap $\Delta\approx\SI{10}{meV}$ extracted from thermal activation measurements. This gap  occurs when no external electric field is applied. 
	Along the zero-density lines for the individual bilayer systems the gap opens and closes as a function of the applied displacement field $D$. Applying $\DT=\SI{-0.12}{V/nm}$ closes the gap in the upper,  $\DB=\SI{0.13}{V/nm}$ closes the gap in the lower layer. We attribute the intrinsic gaps to the existence of crystal fields ($\cfieldB,\cfieldT$) which we have to counteract by applying an external electrical field to close the gaps. 
	
	Due to the finite size of the top gate we are able to define electrostatic in-plane Fabry-Pérot cavities in the top and/or the bottom layer \cite{Rickhaus2019}.  The observed Fabry-Pérot oscillations testify to the fact that the tunnel coupling between the BG layers is much weaker than the in-plane hopping between neighboring atoms, and that the two layers exhibit coherent in-plane transport. We find that the interference pattern of the top layer in the $\VT,\VB$-plane  changes slope if the bottom layer is gapped. The slope in the gap is in agreement with the electrostatic conditions for BG (instead of TDBG) since the gapped layer does not screen electric fields. This demonstrates that transport can be switched on or off by gating in each layer. 
	
	Our experimental results are in agreement with first principles calculations, which predict a gapped band structure with an excess of electrons in the inner layers 
	due to the intrinsic crystal field effect. We show that such an effect can be easily included in low-energy models by an effective crystal field term. Our results demonstrate that these crystal fields will be generically relevant for the low energy dispersion of any van-der-Waals heterostructure with more than 3 layers or an asymmetric vertical arrangement of layers. Combining the experimentally measured band gap and our low energy description, we highlight that crystal fields will also substantially modify the electronic structure at generic twist-angles, and in particular in small-angle twisted double bilayer graphene \cite{Choi2019,PhysRevB.99.235417,PhysRevX.9.031021}. As a result, our work suggests the importance of considering crystal field contributions to faithfully capture the low energy electronic band structure of small-angle twisted double bilayers, where superconductivity and strongly correlated behavior has been reported recently \cite{Liu2019,Shen2019,Burg2019}.
	
	Since the crystal fields can be quite significant ( $\approx0.1-\SI{0.2}{V/nm}$) one may envision to use them for engineering lateral electric field patterns. In the present case, the field arises because the outer graphene layers see hBN on one side and another graphene layer on the other side. Other 2D systems could lead to even larger effective fields across a graphene layer and would enable the combination of built-in electric fields and fields tunable by patterned gate electrodes.

	\section{Methods}
	The TDBG heterostructure is fabricated by first picking up a top hBN flake. With the top hBN we pick up half of a bilayer graphene flake. The stage is then rotated and the other half is picked up \cite{Kim2016,Kim2017}, followed by picking up a bottom hBN flake and a graphite backgate  \cite{Wang2013,Zibrov2017a,Overweg2017a}. The thicknesses of the top and bottom hBN layers $\dT$ and $\dB$ are determined by atomic force microscopy (AFM) and used to calculate the capacitances per area of the TDBG to the top and bottom gates according to $\Ct=\epshBN/\dT$ and $\Cb=\epshBN/\dB$ with $\epsilonhBN=3.3$.
	The structure is contacted by one-dimensional edge-contacts \cite{Wang2013}. Top gates of size $L=\SI{400}{nm}$ in transport direction for devices 1 and 2 are defined by electron-beam lithography. Device 3 exhibits a global top-gate, which is insulated from the contacts by a $\SI{30}{nm}$ thick AlO$_x$ layer.
	Two-terminal linear conductance measurements are performed using a low-frequency lock-in technique ($\SI{177}{Hz}$) at the temperature $T=\SI{2}{K}$.

	\begin{figure}
		\centering
		\includegraphics[width=1\textwidth]{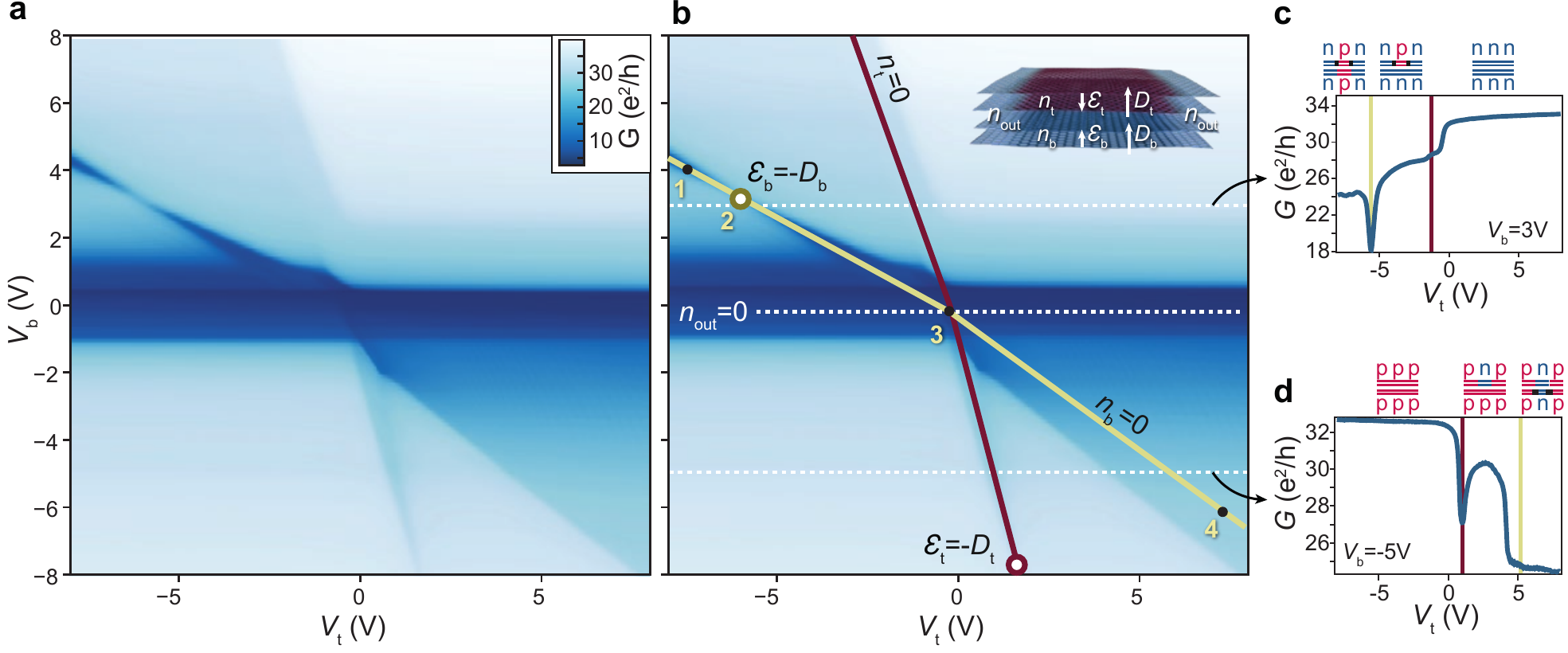}
		\caption{a) Measured conductance as a function of global bottom and local top gate, $G(\VB,\VT)$. 
			b) Same data as in a) with zero density lines in the top- and bottom layer ($\nt=0$, red and $\nb=0$, yellow). Points where the crystal fields are compensated by the external fields $\DT$ and $\DB$ are labeled with $\mathcal{E}_\mathrm{t}=-\DT$ and $\mathcal{E}_\mathrm{b}=-\DB$.
			c) $G(\VT)$ for $\VB=\SI{3}{V}$ and d) $\VB=\SI{-5}{V}$. The observed values for $G$ can be qualitatively understood by considering p-n-p junctions between the double-gated region below the top gate (with densities $\nt$ and $\nb$) and the single-gated regions, which are tuned only by the back gate (with density $\nout$). The appearance of p-n-p junctions in the top,  bottom or both BG leads to a reduction of $G$.
		}\label{fig:2}
	\end{figure}
	
	\subsection{Zero density lines}
	The conductance $G(\VT,\VB)$ as a function of the local top and global backgate of device 1 is shown in Fig. \ref{fig:2}a. We first discuss the lines of zero-density ($\nt=0$ and $\nb=0$ in Fig.\ref{fig:2}b).
	In the double-gated region, changing $\VT$ induces charge carriers in the top BG, which in turn have a gating effect on the bottom BG. Therefore, to evaluate the conditions for zero density we not only need to take into account the capacitance to the top- and back gate ($\Ct$, $\Cb$) but also the geometric capacitance between the top and bottom BG ($\Cm$) and the quantum capacitance of each BG ($\Cqt$, $\Cqb$). The slopes of the zero density lines in the top and bottom BG are then given by \cite{Rickhaus2019}
	\begin{eqnarray*}
		\left.\frac{\partial\VT}{\partial\VB}\right|_{\nb=0}  & \approx & -\frac{\Cb}{\Ct}\left(1+\frac{\Cqt}{\Cm}\right)\\
		\left.\frac{\partial\VB}{\partial\VT}\right|_{\nt=0}  & \approx & -\frac{\Ct}{\Cb}\left(1 + \frac{\Cqb}{\Cm}\right).
	\end{eqnarray*} 
	Here we neglected terms $\Cb/\Cm$ and $\Ct/\Cm$ which are $\ll1$ (for details see Supplemental Material). In the $(\VT,\VB)$-map, a line with slope $-\Ct/\Cb$ corresponds to the gating condition where the total charge carrier density vanishes. In order to reach zero density in e.g.\ the bottom layer, screening has to be taken into account, and it modifies the $\ntot=0$ slope by the factor $(1+\frac{\Cqt}{\Cm})$. We depict calculated zero density lines in Fig.\ref{fig:2}b. 
	
	From our measurement we find $\Ct/\Cb\approx1.6\pm0.2$ which is in agreement with the expected slope considering the measured thickness of top and bottom hBN, i.e.\ $\Ct/\Cb=\dB/\dT=1.5$. For the geometric capacitance between the BG layers we obtain $\Cm=\SI[separate-uncertainty = true]{3.5\pm 1}{\mu F/cm^2}$. We note that $\Cm$ is, within the error bars, half the capacitance between two single-layer graphene sheets ($C_\mathrm{m,SLG}=\SI[separate-uncertainty = true]{7.5\pm 0.7}{\mu F/cm^2}$ \cite{Rickhaus2019}), as expected. Details are given in the Supplemental Material.
	
	Analyzing the zero density lines we extract the quantum capacitance $\Cq$ and thereby the system's density of states. If two single-layer graphene sheets are twisted by a large angle, the measured zero-density lines are curved since $\Cq\propto\sqrt{n}$ \cite{Rickhaus2019}. Here, our system consists of two BG layers, therefore $\Cq=\rm{const.}$, and straight lines are observed. Since $\Cq$ depends on the effective mass, it indicates electron/hole asymmetries of the band structure. We notice that in our measurement the slope of the zero-density line of one BG layer depends on the charge carrier polarity of the other (i.e.\ lines of constant density are kinked at the origin). We interpret this asymmetry as band-dependent screening due to different effective masses $\mh$ and $\me$. Using $\Cqe=2\me e^2/\hbar^2\pi$ and $\Cqh=2\mh e^2/\hbar^2\pi$ we find $\me/\mh=0.63\pm0.1$, i.e., conduction band electrons are lighter than valence band holes. This finding is consistent with experimental results for BG where $\me\approx0.034\melectron$ and $\mh\approx0.044\melectron$ at a charge carrier density of $\SI{2e12}{cm^{-2}}$, i.e.\ $\me/\mh=0.75$ \cite{Li2016}. We show later in the paper that it also agrees qualitatively with DFT calculations for TDBG ($\me/\mh=0.75$) and BG ($\me/\mh=0.77$).
	Residual doping does not play a major role for the interpretation of the experimental data. 
	
	\subsection{Electrical field induced gaps}
	In BG, an electric field $D$ normal to the layers induces a single-particle gap. In double-gated transport experiments \cite{Oostinga2007, Overweg2017a}, this gap leads to a region of low conductance along the zero-density line in a $(\VT,\VB)$-map, along which $D$ changes. The conductance is highest at $D=0$, which is usually close to the origin of the map. The full width half maximum of the low-conductance region along the zero-density line increases with increasing $|D|$, which is attributed to the presence of localized states in the gap.
	
	We observe similar behavior along zero density lines in our TDBG structure. We first focus on the $\nb=0$ line for $\VB>0$. The conductance is highest at $(\VT,\VB)\approx(\SI{-5.8}{V},\SI{3.1}{V})$, and a region of increasing width and decreasing conductance appears with increasing distance from this point along the $\nb=0$ line. We interpret the point $(\VT,\VB)\approx(\SI{-5.8}{V},\SI{3.1}{V})$ as the point of zero displacement field in the bottom layer and attribute the decreasing conductance to the opening of a gap in the bottom BG. We confirm this interpretation  later by analyzing thermal activation and screening properties. Importantly, the gap is closed far off the origin of the entire map, even if residual doping is taken into account. This means that a significant external electrical field needs to be applied to the bottom BG in order to compensate for the finite \emph{internal} field present at zero external field. This internal field is the crystal field in the TDBG as we confirm later with density functional theory (DFT) calculations.
	
	In contrast, along the zero density line $\nt=0$ of the top layer, the gap closes at $(\SI{1.7}{V},\SI{-8}{V})$. For this layer, the gap closes by applying an external electric field with a sign opposite to that needed for the bottom layer. Using the capacitance model we extract the external fields $\DT$ and $\DB$ needed to close the gap in the respective layer (details are given in the Supplemental Material). Thereby we directly measure the internal crystal fields  $\cfieldT=-\DT=\SI{0.12}{V/nm}$ and $\cfieldB=-\DB=\SI{-0.13}{V/nm}$. We find $\cfieldT=-\cfieldB$ within experimental error. 
	
	For device 2 (and device 3) we obtain similar results, i.e.  $\cfieldT=\SI{0.11}{V/nm}$ ($\cfieldT=\SI{0.13}{V/nm}$) and $\cfieldB=\SI{-0.12}{V/nm}$ ($\cfieldB=\SI{-0.14}{V/nm}$). Details are given in the Supplemental Material.
	Having determined the crystal fields, we estimate the size of the crystal-field induced gaps \cite{McCann2006} obtaining a value in the range $\Delta\approx20-\SI{30}{meV}$.

	\subsection{Conductance values}
	\begin{figure}
		\centering
		\includegraphics[width=1\textwidth]{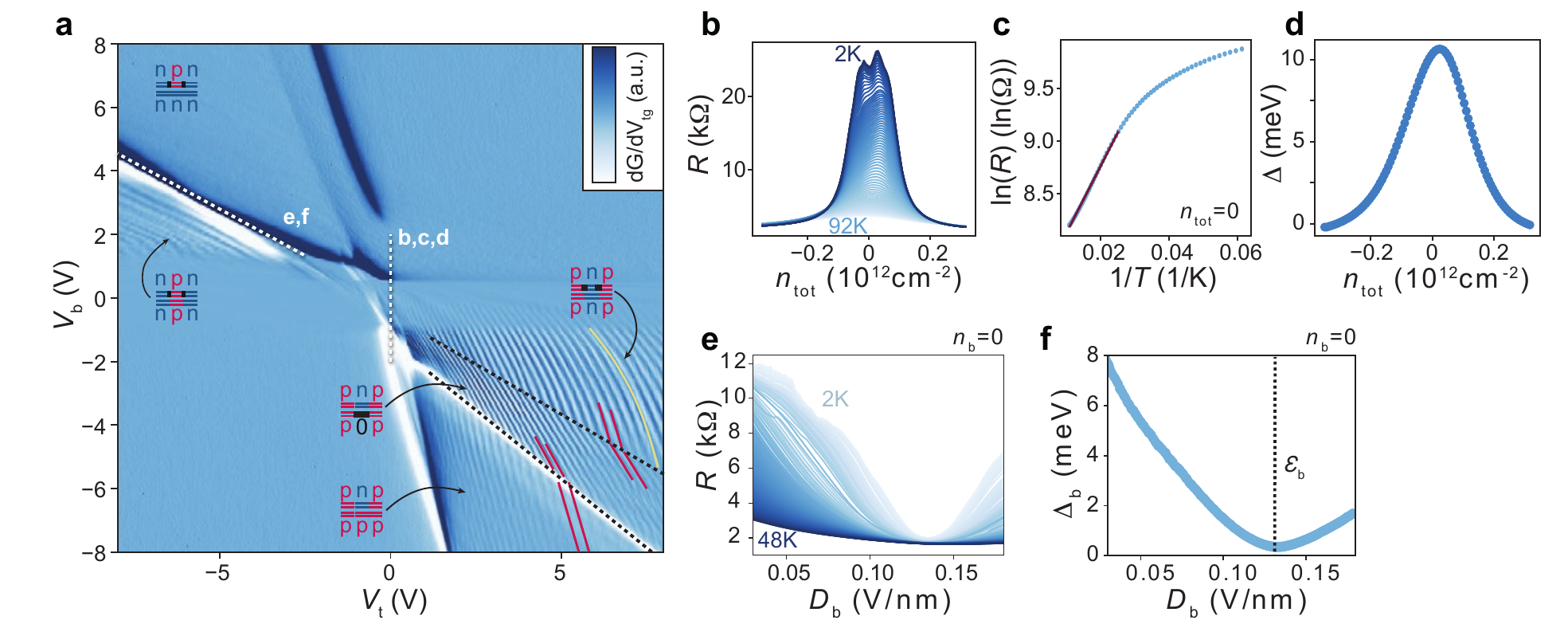}
		\caption{
			a) Numerical derivative $dG/d\VT$ of the conductance map shown in Fig.\ref{fig:2}a (device 1). Fabry-Pérot resonances (red lines) change slope when entering the gapped region in the bottom layer (between the black dashed lines). 
			b) For device 3 with a global topgate, we depict $R(\ntot)$ for temperatures between $\SI{2}{K}$ and $\SI{92}{K}$. The resistance follows a thermally activated behavior. Data of device 1 is shown in the Supplemental Material).
			c) In an Arrhenius plot ($\ntot=0$), the slope at high temperatures (red line) corresponds to a gap of $\Delta=\SI{10.6}{meV}$. In d) we plot the extracted $\Delta(\ntot)$. 
			e) $R(\DB)$ for different temperatures for device 1. The values are taken for a gating configuration along the $\nb=0$ line which is marked with a white dashed line in a). This allows to extract the gap size in the bottom layer $\Delta_b(\DB)$, as shown in f).
		}\label{fig:3}
	\end{figure}
	The finite size of the top gate matters for the conductance since its value is affected by the presence or absence of p-n junctions. The regions in the graphene flake not below the top gate are affected by the back gate only (single-gated regions). Its voltage $\VB$ tunes the single-gated regions to be p- or n-doped. The line of constant density in the single-gated regions $\nout=0$ is a horizontal line in Fig.\ref{fig:2}b. p-n junctions in the top (bottom) BG are formed when crossing $\nt=0$ ($\nb=0$). 
	
	For device 3, which has a global topgate, this line is not present (see Supplemental Material).
	
	Two line cuts $G(\VT)$ for $\VB=3V$ and $\VB=-5V$ are shown in Fig.\ref{fig:2}c and d.
	For $\VT,\,\VB >0$, the single- and double-gated regions are n-doped and the conductance is large. By reducing $\VT$, the line $\nt=0$ is crossed, therefore a p-n-p junction in the top-layer appears and reduces $G$. A further reduction is observed when crossing $\nb=0$, where a p-n-p junction forms in the bottom layer. The inverted situation with n-doped single-gated regions is shown in Fig. \ref{fig:2}d.
	
	The data also reveal that the conductance is affected more strongly when crossing a zero density line with a large gap ($\nt=0$ for $\VB>0$ and $\nb=0$ for $\VB<0$)  than in the opposite case. We attribute this to the presence of a spatially more extended gapped region at the p-n interface if the gap is large.

	\subsection{Fabry-Pérot oscillations}
	The presence of regular Fabry-Pérot resonances is revealed when we look at the numerical derivative $dG/d\VT$ (Fig. \ref{fig:3}a) \cite{Young2009,Rickhaus2013,Varlet2014}. The oscillations occur if a p-n-p or an n-p-n junction is present in either the top- or the bottom BG, see also \cite{Rickhaus2019}. We observe two sets of Fabry-Pérot resonances. For negative $\VB$, resonances parallel to the zero density line in the top layer ($\nt=0$) are observed due to the formation of a p-n-p junction in the upper layer. These resonances are not present for $\VB>0$, i.e.\ if an n-p-n junction is formed. The mirrored situation is observed for resonances in the bottom layer, i.e.\ they are clearly visible in the n-p-n, but not in the p-n-p regime. Also this finding can be attributed to the presence of a spatially extended gapped region at  the p-n interface in the n-p-n (p-n-p) regime for the bottom (top) layer that suppresses the relative contribution of mesoscopic effects of the bottom (top) layer to the total current.
	
	We now focus on the Fabry-Pérot oscillations in the top layer for $\VB<0$ few of which are marked with red lines in Fig. \ref{fig:3}a. Apparently, the slope of the resonances changes if the bottom layer becomes gapped (the region is marked with black dashed lines in the figure). The electrostatics in this situation are reduced to that of a simple bilayer system, where lines of constant density are given by $\left.\partial\VB/\partial\VT\right|_{\ntot=\text{const.}}  \approx  -\Ct/\Cb$. The measured slope is in agreement with this expectation (see Supplemental Material). 	
	Such a change in the electrostatic screening clearly demonstrates that the bottom layer is indeed gapped around $\nb=0$.
	A model for the expected Fabry-Pérot oscillation periodicity, given the layer densities and the length $L=\SI{400}{nm}$ of the topgate is given in the Supplemental material. We find a good agreement with the observed periodicity. The model does not capture the change of slope when the resonances approach $\VB=0$ (yellow line). This can be attributed to an increasing cavity size as the asymmetry of the gating increases \cite{Rickhaus2013,Handschin2017}.
	
	\subsection{Thermal Activation}
	
	We further studied thermal activation of carriers across the gap. In Fig.\ref{fig:3}b we show the resistance as a function of $\ntot$ for temperatures $T$ between $\SI{2}{K}$ and $\SI{92}{K}$. The data is taken on device 3 which has a global topgate. Similar data on device 1 is shown in the Supplemental Material. The resistance shows activated behavior with increasing $T$. Fitting an Arrhenius law ($R_{\mathrm{max}}\propto\exp(\Delta/2k_\mathrm{B}T)$)  we extract a gap of $\Delta=\SI{10.6}{meV}$ at zero gate voltage (see Fig. \ref{fig:3}cd) and $\Delta=\SI{15}{meV}$ for device 1. These values are of the same order of magnitude as the gap estimated before using the size of the measured crystal-field (i.e.\ $\Delta\approx20-\SI{30}{meV}$). Deviations are possibly due to thermally activated hopping through states in the gap.
	\\
	Finally, we measured the opening and closing of the gap in the bottom layer $\Delta_{\rm{b}}$ along the $\nb=0$ line (Fig. \ref{fig:3}ef) for device 1. With increasing external field $\DB$, the gap decreases until it reaches a minimum at $\DB=\cfieldB=\SI{0.13}{V/nm}$. We find that $\Delta_{\rm{b}}(\DB)$ is almost linear, as expected for an electrostatically induced gap in BG. Gate maps at different temperatures can be found in the Supplemental Material.

	\subsection{Theory}
	
	\begin{figure}
		\centering
		\includegraphics[width=1\textwidth]{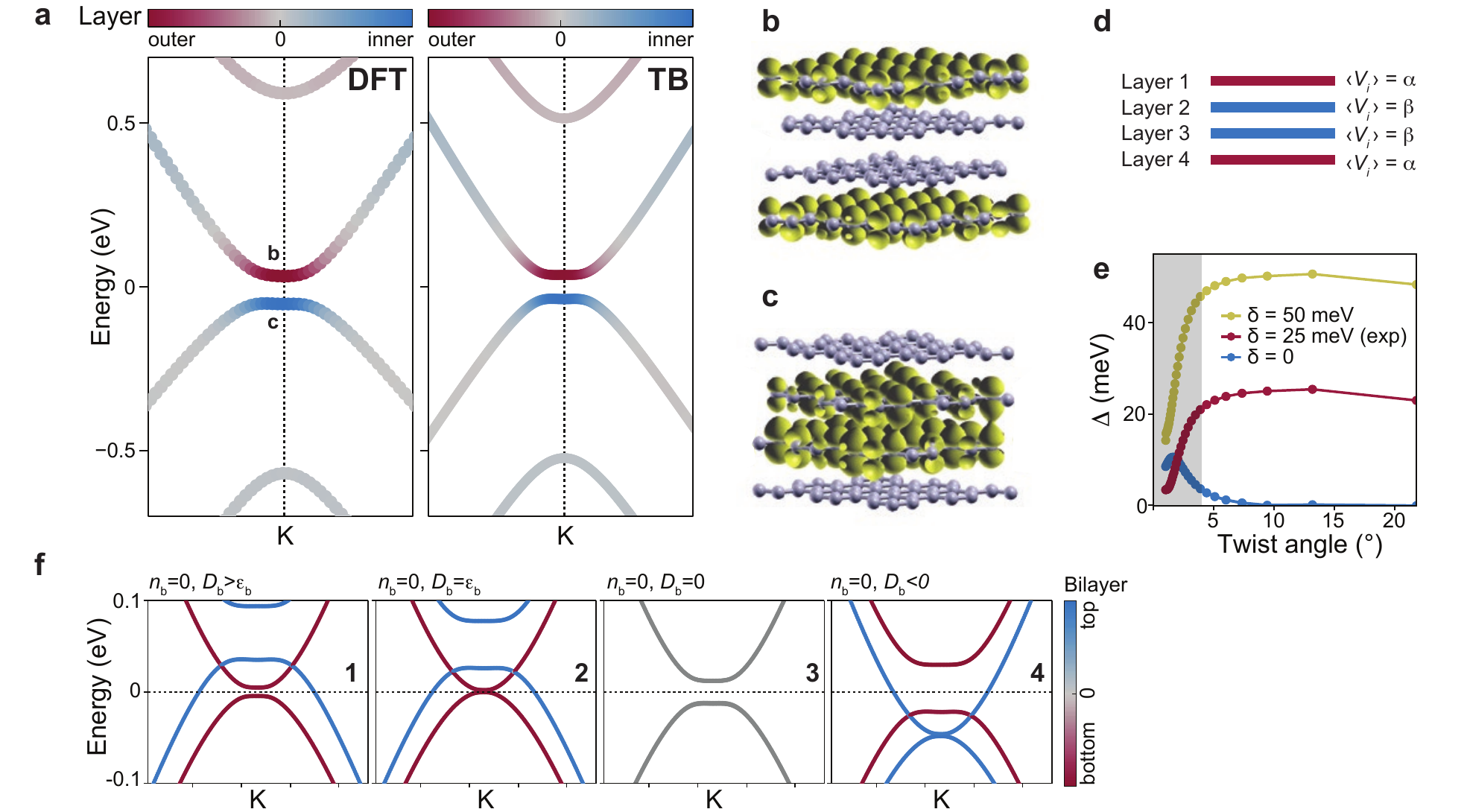}
		\caption{a) Band structure of TDBG with a rotation angle of $\theta=\SI{13}{\degree}$ with density functional theory (left) and a low energy tight binding model (right). 
			The color denotes in which layers (outer or inner) the wavefunction associated with the eigenvalue is localized.
			b) Density isosurface (yellow surface) associated with the conduction band wavefunctions at the K-point for $\theta=\SI{13}{\degree}$, showing a state localized in the two outer graphene layers, as obtained from the DFT calculation.
			c) For the valence band wavefunctions, the state is localized in the inner graphene layers.
			d) In the tight binding model, the crystal field contribution is effectively captured by the layer dependent on-site energy $\langle V_i\rangle$. For the inner layers $\langle V_i\rangle=\beta$ and the outer layers $\langle V_i\rangle=\alpha$ with $\alpha\neq\beta$.
			e) Gap at the K point as a function of the twist angle, for various values of the crystal field contribution. It is observed that at large angles the gap scales
			linearly with the crystal field layer imbalance $\delta$, whereas at low angles the crystal field gap competes with the intrinsic band gap.
			f) Band structure of TDBG with the crystal field contribution as function of an external interlayer bias, showing that the relative gaps in the two layers can be independently
			controlled. The numbers 1-4 correspond to different gating configurations, labelled in Fig.\ref{fig:2}b.
		}\label{fig:4}
	\end{figure}
	
	To asses the single particle nature of the experimentally observed band gap, we have performed first principles calculations of TDBG graphene, with a rotation angle between the layers of $13$ degrees. We show in Fig. \ref{fig:4}a the first principles band structure, showing a band gap at the K point of the mini-Brillouin zone. To understand the origin of such a band gap, we computed the layer distribution of the associated Bloch wave functions, to distinguish between states located on the two inner layers and the two outer layers. We do so by defining the inner-outer operator $\Theta$ as
	\begin{equation}
	\Theta = P_2 + P_3 - P_1 - P_4
	\end{equation}
	where $P_i$ is the projection operator in the atomic orbitals of layer $i$. Computing the expectation value $\langle \Psi | \Theta | \Psi \rangle$ for each Bloch wave function $\Psi$ allows us to identify the spatial location of each individual state. As shown in Fig.  \ref{fig:4}a,  the inner-outer polarization $\Theta$
	shows that the top of the valence band hosts states strongly localized in the two inner layers, whereas the bottom of the conduction band hosts states mainly localized in the two outer layers. 
	That picture can be further confirmed by looking at the spatial distribution
	of the wave functions at the $K$ point (Fig. \ref{fig:4}bc), showing that the valence state is localized in the two inner layers, whereas the conduction state is localized in the two outer layers.
	Such layer imbalance creates a small charge imbalance between the two layers, yielding negatively charged inner layers together with positively charged outer layers at charge neutrality, in the absence of any external bias. We emphasize that this is a purely intrinsic electrostatic effect,
	and it is not associated with an interaction induced symmetry broken state, as it simply stems from the inequivalence of the inner/outer layers in the structure.
	
	It is important to note that the first principles calculations were performed for a free-standing double bilayer, whereas in the experiment the double bilayer is encapsulated between hBN. The hBN layers will thus create additional crystal fields in the outer layer, modifying the net internal field of the whole system. 
	By performing a calculation hosting simultaneously a graphene monolayer encapsulated in hBN and a free standing one, we have estimated the crystal field induced by the hBN. It was found that proximity to hBN lowers the onsite energies of the carbon atoms in comparison with the free standing case. As a result, the hBN in the twisted double bilayer system will slightly counteract the intrinsic crystal field, reducing the effective internal field and thus reducing the value of the band gap (see Supplemental Material). As a result, a free standing twisted double bilayer graphene is expected to show a larger intrinsic gap than the encapsulated one discussed in this manuscript.
	
	Both the layer imbalance and the gap at the $K$ point does not appear in low energy models of TDBG based on a continuum Dirac equation or tight binding model.
	Such absence stems from the electrostatic origin of the phenomenon. 
	We have incorporated the effect of this electrostatic crystal field effect in a low energy tight binding model of the form
	
	\begin{equation}
	H_0 = 
	-t\sum_{\langle ij \rangle}
	c^\dagger_i c_j
	+
	\sum_{ij}
	t_\perp (\bold r_i,\bold r_j) c^\dagger_i c_j
	+
	\sum_i V(\bold r_i) c^\dagger_i c_i
	-\mu \sum_{i} c^\dagger_i c_i
	\end{equation}
	where $\langle ij \rangle$ denotes the sum over first neighbors in the same layer, $\bold r_i$ are the different positions of the carbon atoms in the TDBG structure, and $\mu$ is an overall chemical shift chosen to have the charge neutrality point at $E=0$.
	We use the interlayer hopping function
	$t_{\perp}(\bold r_i,\bold r_j) = 
	t_{\perp}
	\frac{(z_i - z_j)^2 }{|\bold r_i - \bold r_j|^2}
	e^{-\beta (|\bold r_i - \bold r_j|-d)},
	$
	where $\bold r_i$ is the real space position of carbon atom with index $i$
	and $d$ the distance between layers. We take the hopping parameters 
	$t =2.7 $ eV
	and $t_\perp = 0.4$ eV.

	The function $V(\bold r_i)$ denotes the crystal field function that we take to be of the form
	\begin{equation}
	V(\bold r_i) = \lambda \sum_{j\ne i} \frac {e^{-|\bold r_i-\bold r_j|/\Lambda}}{|\bold r_i- \bold r_j|}
	\end{equation}
	where $\lambda$ controls the strength of the crystal field and $\Lambda$ is the decay constant of the crystal field. The previous functional form can be readily used to include crystal field effects in an arbitrary twisted graphene multilayer. Analogous crystal field terms have long been known to be present in graphite, and are a natural consequence of the overlap of electronic clouds between different carbon atoms \cite{PhysRevB.15.4077,PhysRevB.78.205425}.
	For single layer graphene, this term simply adds an overall constant
	to the Hamiltonian, and therefore does not affect the electronic spectra. 
	In a twisted graphene bilayer, this term creates a small modulation in the Moiré unit cell  analogous to a Hartree term of the Coulomb interaction \cite{Guinea2018}, yet without opening any gap in the single particle spectra. For multilayer van der Waals heterostructures with more than two layers\cite{PhysRevB.98.045407}, and in particular for the twisted double bilayer described in this manuscript,
	this term will make inner and outer layers inequivalent by inducing
	different electrostatic potentials, effectively creating an intrinsic interlayer bias \cite{2019arXiv190600623H}.
	In this TDBG, at large angles the electronic states of each layer
	are decoupled at low energies, and the intrinsic interlayer bias in each bilayer opens up a gap \cite{Castro2007,2019arXiv190600623H}.
	
	The effect of the crystal field term in the TDBG can be characterized by the average difference between the onsite energies between the inner and outer layers $\delta$ defined as
	
	\begin{equation}
	\delta = \alpha - \beta =
	\langle V \rangle_{\text{Layer } 1} -
	\langle V \rangle_{\text{Layer } 2} 
	\end{equation}
	where $ \langle \rangle_{\text{Layer } j}$ denotes the average value over atoms in layer $j$. The potential difference $\delta$ is controlled in the tight binding model mainly by the parameter
	$\lambda$, with a weak dependence on the screening length $\Lambda$. Importantly, since the effective crystal field can depend on the materials between which the TDBG is encapsulated, the linear dependence of the gap allows one to extract the crystal field parameters for the effective tight binding model by comparing with the experimentally measured intrinsic band gap. As noted before, such intrinsic band gap is expected to be different for free-standing bi-bilayer and the encapsulated one of this work, and therefore the crystal field $\delta$ will depend on the encapsulation of the bi-bilayer.
	
	The combined effect of crystal field and external electric bias can be included in the tight binding model. In Fig. \ref{fig:4}f we show the band structure of the TDBG including the crystal field contribution and an external bias perpendicular to the bi-bilayer. The numbers 1-4 in the figure correspond to gating configurations which are labeled in Fig. \ref{fig:2}b. As the interlayer bias is ramped from negative to positive values, the gaps in the two layers change independently due to the interplay between crystal field and interlayer bias. The effective decoupling between the two layers is valid in the present large angle regime, whereas at smaller angles the low energy states will be hybridized between the two layers.
	
	The impact of the crystal field contribution at a twist angle of 13 degrees suggests that its effect can introduce important modifications at smaller angles, where the crystal field gap may compete with the intrinsic band gap of small-angle TDBG that stems purely from the interlayer hopping \cite{2019arXiv190300852C,PhysRevB.99.235417,PhysRevB.99.235406}. To address this problem, we exploit the tight-binding model to compute the band gap at the K-point as a function of the twist between the two bilayers with the crystal field contribution, as shown in Fig. \ref{fig:4}e. In the absence of the crystal field contribution, the bilayer shows a negligible gap at large angle, which becomes only sizable at small angles. In contrast, in the presence of the crystal field term, the bilayer shows a  generic band gap for large angles, that decreases when the small angle regime is approached. At small angles (shaded area in Fig. \ref{fig:4}e) additional lattice relaxation effects may come into play that are not captured by our tight binding model. The behavior of the gap as a function of the angle is qualitatively the same for different values of the crystal field, suggesting that both the encapsulated TDBG of this work and a free standing one would show analogous phenomenology.

	\subsection{Conclusion}
	We have shown experimentally that pristine twisted double bilayer graphene with a large twist angle between the bilayers has an intrinsic band gap in each layer, that can be closed by applying an external interlayer bias. The origin of the intrinsic band gap, which is a single particle effect, lies in the electrostatics of the different graphene layers. The inequivalency between the inner and outer layers results in different potentials for electrons leading to a small electron transfer, which opens a band gap. This scenario has been verified both with first principles calculations and a low-energy model. Both capture the existence of the intrinsic band gap, as well as its selective closing by means of an external interlayer bias.
	
	The mechanism opening the gap is in striking contrast to other scenarios studying both aligned and twisted graphene layers. There, the intrinsic bandgaps are associated with many-body effects. Our results prove that crystal field effects can have an important impact on the electronic structure of twisted double bilayers, and must be included in theoretical models to properly account for the physics of twisted double layer graphene both at large and small angles. The ability to open and close the gap in each of the bilayers individually may be useful for exploiting the layer degree of freedom in quantum computation applications.

	\section*{Acknowledgements}
	We acknowledge financial support from the European Graphene Flagship, the Swiss National Science Foundation via NCCR Quantum Science. J. L. Lado and P. Rickhaus acknowledge financial support from the ETH Fellowship program. R. Garreis acknowledges funding from the European Union’s Horizon 2020 research and innovation programme under the Marie Skłodowska-Curie grant agreement No 766025. Growth of hexagonal boron nitride crystals was supported by the Elemental Strategy Initiative conducted by MEXT, Japan and the CREST (JPMJCR15F3), JST.

	\bibliography{tBBG}

\begin{thebibliography}{42}%
\makeatletter
\providecommand \@ifxundefined [1]{%
 \@ifx{#1\undefined}
}%
\providecommand \@ifnum [1]{%
 \ifnum #1\expandafter \@firstoftwo
 \else \expandafter \@secondoftwo
 \fi
}%
\providecommand \@ifx [1]{%
 \ifx #1\expandafter \@firstoftwo
 \else \expandafter \@secondoftwo
 \fi
}%
\providecommand \natexlab [1]{#1}%
\providecommand \enquote  [1]{``#1''}%
\providecommand \bibnamefont  [1]{#1}%
\providecommand \bibfnamefont [1]{#1}%
\providecommand \citenamefont [1]{#1}%
\providecommand \href@noop [0]{\@secondoftwo}%
\providecommand \href [0]{\begingroup \@sanitize@url \@href}%
\providecommand \@href[1]{\@@startlink{#1}\@@href}%
\providecommand \@@href[1]{\endgroup#1\@@endlink}%
\providecommand \@sanitize@url [0]{\catcode `\\12\catcode `\$12\catcode
  `\&12\catcode `\#12\catcode `\^12\catcode `\_12\catcode `\%12\relax}%
\providecommand \@@startlink[1]{}%
\providecommand \@@endlink[0]{}%
\providecommand \url  [0]{\begingroup\@sanitize@url \@url }%
\providecommand \@url [1]{\endgroup\@href {#1}{\urlprefix }}%
\providecommand \urlprefix  [0]{URL }%
\providecommand \Eprint [0]{\href }%
\providecommand \doibase [0]{http://dx.doi.org/}%
\providecommand \selectlanguage [0]{\@gobble}%
\providecommand \bibinfo  [0]{\@secondoftwo}%
\providecommand \bibfield  [0]{\@secondoftwo}%
\providecommand \translation [1]{[#1]}%
\providecommand \BibitemOpen [0]{}%
\providecommand \bibitemStop [0]{}%
\providecommand \bibitemNoStop [0]{.\EOS\space}%
\providecommand \EOS [0]{\spacefactor3000\relax}%
\providecommand \BibitemShut  [1]{\csname bibitem#1\endcsname}%
\let\auto@bib@innerbib\@empty
\bibitem [{\citenamefont {Gr{\"{u}}neis}\ \emph {et~al.}(2008)\citenamefont
  {Gr{\"{u}}neis}, \citenamefont {Attaccalite}, \citenamefont {Wirtz},
  \citenamefont {Shiozawa}, \citenamefont {Saito}, \citenamefont {Pichler},\
  and\ \citenamefont {Rubio}}]{Gruneis2008}%
  \BibitemOpen
  \bibfield  {author} {\bibinfo {author} {\bibfnamefont {A.}~\bibnamefont
  {Gr{\"{u}}neis}}, \bibinfo {author} {\bibfnamefont {C.}~\bibnamefont
  {Attaccalite}}, \bibinfo {author} {\bibfnamefont {L.}~\bibnamefont {Wirtz}},
  \bibinfo {author} {\bibfnamefont {H.}~\bibnamefont {Shiozawa}}, \bibinfo
  {author} {\bibfnamefont {R.}~\bibnamefont {Saito}}, \bibinfo {author}
  {\bibfnamefont {T.}~\bibnamefont {Pichler}}, \ and\ \bibinfo {author}
  {\bibfnamefont {A.}~\bibnamefont {Rubio}},\ }\bibfield  {title} {\enquote
  {\bibinfo {title} {{Tight-binding description of the quasiparticle dispersion
  of graphite and few-layer graphene}},}\ }\href {\doibase
  10.1103/PhysRevB.78.205425} {\bibfield  {journal} {\bibinfo  {journal} {Phys.
  Rev. B - Condens. Matter Mater. Phys.}\ }\textbf {\bibinfo {volume} {78}},\
  \bibinfo {pages} {205425} (\bibinfo {year} {2008})}\BibitemShut {NoStop}%
\bibitem [{\citenamefont {{Haddadi}}\ \emph {et~al.}(2019)\citenamefont
  {{Haddadi}}, \citenamefont {{Wu}}, \citenamefont {{Kruchkov}},\ and\
  \citenamefont {{Yazyev}}}]{2019arXiv190600623H}%
  \BibitemOpen
  \bibfield  {author} {\bibinfo {author} {\bibfnamefont {Fatemeh}\ \bibnamefont
  {{Haddadi}}}, \bibinfo {author} {\bibfnamefont {QuanSheng}\ \bibnamefont
  {{Wu}}}, \bibinfo {author} {\bibfnamefont {Alex~J.}\ \bibnamefont
  {{Kruchkov}}}, \ and\ \bibinfo {author} {\bibfnamefont {Oleg~V.}\
  \bibnamefont {{Yazyev}}},\ }\bibfield  {title} {\enquote {\bibinfo {title}
  {{Moir\textbackslash'e Flat Bands in Twisted Double Bilayer Graphene}},}\
  }\href@noop {} {\bibfield  {journal} {\bibinfo  {journal} {arXiv e-prints}\
  ,\ \bibinfo {eid} {arXiv:1906.00623}} (\bibinfo {year} {2019})},\ \Eprint
  {http://arxiv.org/abs/1906.00623} {arXiv:1906.00623 [cond-mat.mes-hall]}
  \BibitemShut {NoStop}%
\bibitem [{\citenamefont {Rozhkov}\ \emph {et~al.}(2016)\citenamefont
  {Rozhkov}, \citenamefont {Sboychakov}, \citenamefont {Rakhmanov},\ and\
  \citenamefont {Nori}}]{Rozhkov2016}%
  \BibitemOpen
  \bibfield  {author} {\bibinfo {author} {\bibfnamefont {A.~V.}\ \bibnamefont
  {Rozhkov}}, \bibinfo {author} {\bibfnamefont {A.~O.}\ \bibnamefont
  {Sboychakov}}, \bibinfo {author} {\bibfnamefont {A.~L.}\ \bibnamefont
  {Rakhmanov}}, \ and\ \bibinfo {author} {\bibfnamefont {Franco}\ \bibnamefont
  {Nori}},\ }\bibfield  {title} {\enquote {\bibinfo {title} {{Electronic
  properties of graphene-based bilayer systems}},}\ }\href {\doibase
  10.1016/j.physrep.2016.07.003} {\bibfield  {journal} {\bibinfo  {journal}
  {Phys. Rep.}\ }\textbf {\bibinfo {volume} {648}} (\bibinfo {year} {2016}),\
  10.1016/j.physrep.2016.07.003},\ \Eprint {http://arxiv.org/abs/1511.06706}
  {arXiv:1511.06706} \BibitemShut {NoStop}%
\bibitem [{\citenamefont {Luican}\ \emph {et~al.}(2011)\citenamefont {Luican},
  \citenamefont {Li}, \citenamefont {Reina}, \citenamefont {Kong},
  \citenamefont {Nair}, \citenamefont {Novoselov}, \citenamefont {Geim},\ and\
  \citenamefont {Andrei}}]{Lucian2011}%
  \BibitemOpen
  \bibfield  {author} {\bibinfo {author} {\bibfnamefont {A}~\bibnamefont
  {Luican}}, \bibinfo {author} {\bibfnamefont {Guohong}\ \bibnamefont {Li}},
  \bibinfo {author} {\bibfnamefont {A}~\bibnamefont {Reina}}, \bibinfo {author}
  {\bibfnamefont {J}~\bibnamefont {Kong}}, \bibinfo {author} {\bibfnamefont
  {R~R}\ \bibnamefont {Nair}}, \bibinfo {author} {\bibfnamefont {K~S}\
  \bibnamefont {Novoselov}}, \bibinfo {author} {\bibfnamefont {A~K}\
  \bibnamefont {Geim}}, \ and\ \bibinfo {author} {\bibfnamefont {E~Y}\
  \bibnamefont {Andrei}},\ }\bibfield  {title} {\enquote {\bibinfo {title}
  {{Single-Layer Behavior and Its Breakdown in Twisted Graphene Layers}},}\
  }\href {\doibase 10.1103/PhysRevLett.106.126802} {\bibfield  {journal}
  {\bibinfo  {journal} {Phys. Rev. Lett.}\ }\textbf {\bibinfo {volume} {106}},\
  \bibinfo {pages} {126802} (\bibinfo {year} {2011})}\BibitemShut {NoStop}%
\bibitem [{\citenamefont {Sanchez-Yamagishi}\ \emph {et~al.}(2012)\citenamefont
  {Sanchez-Yamagishi}, \citenamefont {Taychatanapat}, \citenamefont {Watanabe},
  \citenamefont {Taniguchi}, \citenamefont {Yacoby},\ and\ \citenamefont
  {Jarillo-Herrero}}]{Sanchez2012}%
  \BibitemOpen
  \bibfield  {author} {\bibinfo {author} {\bibfnamefont {Javier~D}\
  \bibnamefont {Sanchez-Yamagishi}}, \bibinfo {author} {\bibfnamefont {Thiti}\
  \bibnamefont {Taychatanapat}}, \bibinfo {author} {\bibfnamefont {Kenji}\
  \bibnamefont {Watanabe}}, \bibinfo {author} {\bibfnamefont {Takashi}\
  \bibnamefont {Taniguchi}}, \bibinfo {author} {\bibfnamefont {Amir}\
  \bibnamefont {Yacoby}}, \ and\ \bibinfo {author} {\bibfnamefont {Pablo}\
  \bibnamefont {Jarillo-Herrero}},\ }\bibfield  {title} {\enquote {\bibinfo
  {title} {{Quantum Hall Effect, Screening, and Layer-Polarized Insulating
  States in Twisted Bilayer Graphene}},}\ }\href {\doibase
  10.1103/PhysRevLett.108.076601} {\bibfield  {journal} {\bibinfo  {journal}
  {Phys. Rev. Lett.}\ }\textbf {\bibinfo {volume} {108}},\ \bibinfo {pages}
  {76601} (\bibinfo {year} {2012})}\BibitemShut {NoStop}%
\bibitem [{\citenamefont {Rickhaus}\ \emph {et~al.}(2019)\citenamefont
  {Rickhaus}, \citenamefont {Liu}, \citenamefont {Kurpas}, \citenamefont
  {Kurzmann}, \citenamefont {Lee}, \citenamefont {Overweg}, \citenamefont
  {Pisoni}, \citenamefont {Tamaguchi}, \citenamefont {Wantanabe}, \citenamefont
  {Richter}, \citenamefont {Ensslin},\ and\ \citenamefont
  {Ihn}}]{Rickhaus2019}%
  \BibitemOpen
  \bibfield  {author} {\bibinfo {author} {\bibfnamefont {Peter}\ \bibnamefont
  {Rickhaus}}, \bibinfo {author} {\bibfnamefont {Ming-hao}\ \bibnamefont
  {Liu}}, \bibinfo {author} {\bibfnamefont {Marcin}\ \bibnamefont {Kurpas}},
  \bibinfo {author} {\bibfnamefont {Annika}\ \bibnamefont {Kurzmann}}, \bibinfo
  {author} {\bibfnamefont {Yongjin}\ \bibnamefont {Lee}}, \bibinfo {author}
  {\bibfnamefont {Hiske}\ \bibnamefont {Overweg}}, \bibinfo {author}
  {\bibfnamefont {Riccardo}\ \bibnamefont {Pisoni}}, \bibinfo {author}
  {\bibfnamefont {Takashi}\ \bibnamefont {Tamaguchi}}, \bibinfo {author}
  {\bibfnamefont {Kenji}\ \bibnamefont {Wantanabe}}, \bibinfo {author}
  {\bibfnamefont {Klaus}\ \bibnamefont {Richter}}, \bibinfo {author}
  {\bibfnamefont {Klaus}\ \bibnamefont {Ensslin}}, \ and\ \bibinfo {author}
  {\bibfnamefont {Thomas}\ \bibnamefont {Ihn}},\ }\bibfield  {title} {\enquote
  {\bibinfo {title} {{The Electronic Thickness of Graphene}},}\ }\href@noop {}
  {\bibfield  {journal} {\bibinfo  {journal} {arXiv:1907.00582v1}\ ,\ \bibinfo
  {pages} {1--15}} (\bibinfo {year} {2019})},\ \Eprint
  {http://arxiv.org/abs/1907.00582v1} {arXiv:1907.00582v1} \BibitemShut
  {NoStop}%
\bibitem [{\citenamefont {McCann}(2006)}]{McCann2006}%
  \BibitemOpen
  \bibfield  {author} {\bibinfo {author} {\bibfnamefont {Edward}\ \bibnamefont
  {McCann}},\ }\bibfield  {title} {\enquote {\bibinfo {title} {{Asymmetry gap
  in the electronic band structure of bilayer graphene}},}\ }\href {\doibase
  10.1103/PhysRevB.74.161403} {\bibfield  {journal} {\bibinfo  {journal} {Phys.
  Rev. B}\ }\textbf {\bibinfo {volume} {74}},\ \bibinfo {pages} {161403}
  (\bibinfo {year} {2006})}\BibitemShut {NoStop}%
\bibitem [{\citenamefont {Castro}\ \emph {et~al.}(2007)\citenamefont {Castro},
  \citenamefont {Novoselov}, \citenamefont {Morozov}, \citenamefont {Peres},
  \citenamefont {dos Santos}, \citenamefont {Nilsson}, \citenamefont {Guinea},
  \citenamefont {Geim},\ and\ \citenamefont {Neto}}]{Castro2007}%
  \BibitemOpen
  \bibfield  {author} {\bibinfo {author} {\bibfnamefont {Eduardo~V}\
  \bibnamefont {Castro}}, \bibinfo {author} {\bibfnamefont {K~S}\ \bibnamefont
  {Novoselov}}, \bibinfo {author} {\bibfnamefont {S~V}\ \bibnamefont
  {Morozov}}, \bibinfo {author} {\bibfnamefont {N~M~R}\ \bibnamefont {Peres}},
  \bibinfo {author} {\bibfnamefont {J~M B~Lopes}\ \bibnamefont {dos Santos}},
  \bibinfo {author} {\bibfnamefont {Johan}\ \bibnamefont {Nilsson}}, \bibinfo
  {author} {\bibfnamefont {F}~\bibnamefont {Guinea}}, \bibinfo {author}
  {\bibfnamefont {A~K}\ \bibnamefont {Geim}}, \ and\ \bibinfo {author}
  {\bibfnamefont {A~H~Castro}\ \bibnamefont {Neto}},\ }\bibfield  {title}
  {\enquote {\bibinfo {title} {{Biased Bilayer Graphene: Semiconductor with a
  Gap Tunable by the Electric Field Effect}},}\ }\href {\doibase
  10.1103/PhysRevLett.99.216802} {\bibfield  {journal} {\bibinfo  {journal}
  {Phys. Rev. Lett.}\ }\textbf {\bibinfo {volume} {99}},\ \bibinfo {pages}
  {216802} (\bibinfo {year} {2007})}\BibitemShut {NoStop}%
\bibitem [{\citenamefont {Oostinga}\ \emph {et~al.}(2007)\citenamefont
  {Oostinga}, \citenamefont {Heersche}, \citenamefont {Liu}, \citenamefont
  {Morpurgo},\ and\ \citenamefont {Vandersypen}}]{Oostinga2007}%
  \BibitemOpen
  \bibfield  {author} {\bibinfo {author} {\bibfnamefont {Jeroen~B}\
  \bibnamefont {Oostinga}}, \bibinfo {author} {\bibfnamefont {Hubert~B}\
  \bibnamefont {Heersche}}, \bibinfo {author} {\bibfnamefont {Xinglan}\
  \bibnamefont {Liu}}, \bibinfo {author} {\bibfnamefont {Alberto~F}\
  \bibnamefont {Morpurgo}}, \ and\ \bibinfo {author} {\bibfnamefont {Lieven
  M~K}\ \bibnamefont {Vandersypen}},\ }\bibfield  {title} {\enquote {\bibinfo
  {title} {{Gate-induced insulating state in bilayer graphene devices}},}\
  }\href
  {https://www.nature.com/articles/nmat2082{\#}supplementary-information}
  {\bibfield  {journal} {\bibinfo  {journal} {Nat. Mater.}\ }\textbf {\bibinfo
  {volume} {7}},\ \bibinfo {pages} {151} (\bibinfo {year} {2007})}\BibitemShut
  {NoStop}%
\bibitem [{\citenamefont {Zhang}\ \emph {et~al.}(2009)\citenamefont {Zhang},
  \citenamefont {Tang}, \citenamefont {Girit}, \citenamefont {Hao},
  \citenamefont {Martin}, \citenamefont {Zettl}, \citenamefont {Crommie},
  \citenamefont {Shen},\ and\ \citenamefont {Wang}}]{Zhang2009}%
  \BibitemOpen
  \bibfield  {author} {\bibinfo {author} {\bibfnamefont {Yuanbo}\ \bibnamefont
  {Zhang}}, \bibinfo {author} {\bibfnamefont {Tsung-Ta}\ \bibnamefont {Tang}},
  \bibinfo {author} {\bibfnamefont {Caglar}\ \bibnamefont {Girit}}, \bibinfo
  {author} {\bibfnamefont {Zhao}\ \bibnamefont {Hao}}, \bibinfo {author}
  {\bibfnamefont {Michael~C}\ \bibnamefont {Martin}}, \bibinfo {author}
  {\bibfnamefont {Alex}\ \bibnamefont {Zettl}}, \bibinfo {author}
  {\bibfnamefont {Michael~F}\ \bibnamefont {Crommie}}, \bibinfo {author}
  {\bibfnamefont {Y~Ron}\ \bibnamefont {Shen}}, \ and\ \bibinfo {author}
  {\bibfnamefont {Feng}\ \bibnamefont {Wang}},\ }\bibfield  {title} {\enquote
  {\bibinfo {title} {{Direct observation of a widely tunable bandgap in bilayer
  graphene}},}\ }\href
  {https://www.nature.com/articles/nature08105{\#}supplementary-information}
  {\bibfield  {journal} {\bibinfo  {journal} {Nature}\ }\textbf {\bibinfo
  {volume} {459}},\ \bibinfo {pages} {820} (\bibinfo {year}
  {2009})}\BibitemShut {NoStop}%
\bibitem [{\citenamefont {Goossens}\ \emph {et~al.}(2012)\citenamefont
  {Goossens}, \citenamefont {Driessen}, \citenamefont {Baart}, \citenamefont
  {Watanabe}, \citenamefont {Taniguchi},\ and\ \citenamefont
  {Vandersypen}}]{Goossens2012}%
  \BibitemOpen
  \bibfield  {author} {\bibinfo {author} {\bibfnamefont {Augustinus (Stijn)~M}\
  \bibnamefont {Goossens}}, \bibinfo {author} {\bibfnamefont {Stefanie C~M}\
  \bibnamefont {Driessen}}, \bibinfo {author} {\bibfnamefont {Tim~A}\
  \bibnamefont {Baart}}, \bibinfo {author} {\bibfnamefont {Kenji}\ \bibnamefont
  {Watanabe}}, \bibinfo {author} {\bibfnamefont {Takashi}\ \bibnamefont
  {Taniguchi}}, \ and\ \bibinfo {author} {\bibfnamefont {Lieven M~K}\
  \bibnamefont {Vandersypen}},\ }\bibfield  {title} {\enquote {\bibinfo {title}
  {{Gate-Defined Confinement in Bilayer Graphene-Hexagonal Boron Nitride Hybrid
  Devices}},}\ }\href {\doibase 10.1021/nl301986q} {\bibfield  {journal}
  {\bibinfo  {journal} {Nano Lett.}\ }\textbf {\bibinfo {volume} {12}},\
  \bibinfo {pages} {4656--4660} (\bibinfo {year} {2012})}\BibitemShut {NoStop}%
\bibitem [{\citenamefont {Overweg}\ \emph {et~al.}(2017)\citenamefont
  {Overweg}, \citenamefont {Eggimann}, \citenamefont {Chen}, \citenamefont
  {Slizovskiy}, \citenamefont {Eich}, \citenamefont {Pisoni}, \citenamefont
  {Lee}, \citenamefont {Rickhaus}, \citenamefont {Watanabe}, \citenamefont
  {Taniguchi}, \citenamefont {Fal'ko}, \citenamefont {Ihn},\ and\ \citenamefont
  {Ensslin}}]{Overweg2017a}%
  \BibitemOpen
  \bibfield  {author} {\bibinfo {author} {\bibfnamefont {Hiske}\ \bibnamefont
  {Overweg}}, \bibinfo {author} {\bibfnamefont {Hannah}\ \bibnamefont
  {Eggimann}}, \bibinfo {author} {\bibfnamefont {Xi}~\bibnamefont {Chen}},
  \bibinfo {author} {\bibfnamefont {Sergey}\ \bibnamefont {Slizovskiy}},
  \bibinfo {author} {\bibfnamefont {Marius}\ \bibnamefont {Eich}}, \bibinfo
  {author} {\bibfnamefont {Riccardo}\ \bibnamefont {Pisoni}}, \bibinfo {author}
  {\bibfnamefont {Yongjin}\ \bibnamefont {Lee}}, \bibinfo {author}
  {\bibfnamefont {Peter}\ \bibnamefont {Rickhaus}}, \bibinfo {author}
  {\bibfnamefont {Kenji}\ \bibnamefont {Watanabe}}, \bibinfo {author}
  {\bibfnamefont {Takashi}\ \bibnamefont {Taniguchi}}, \bibinfo {author}
  {\bibfnamefont {Vladimir}\ \bibnamefont {Fal'ko}}, \bibinfo {author}
  {\bibfnamefont {Thomas}\ \bibnamefont {Ihn}}, \ and\ \bibinfo {author}
  {\bibfnamefont {Klaus}\ \bibnamefont {Ensslin}},\ }\bibfield  {title}
  {\enquote {\bibinfo {title} {{Electrostatically Induced Quantum Point
  Contacts in Bilayer Graphene}},}\ }\href {\doibase
  10.1021/acs.nanolett.7b04666} {\bibfield  {journal} {\bibinfo  {journal}
  {Nano Lett.}\ }\textbf {\bibinfo {volume} {18}},\ \bibinfo {pages} {553--559}
  (\bibinfo {year} {2017})}\BibitemShut {NoStop}%
\bibitem [{\citenamefont {Allen}\ \emph {et~al.}(2012)\citenamefont {Allen},
  \citenamefont {Martin},\ and\ \citenamefont {Yacoby}}]{Allen2012}%
  \BibitemOpen
  \bibfield  {author} {\bibinfo {author} {\bibfnamefont {M.~T.}\ \bibnamefont
  {Allen}}, \bibinfo {author} {\bibfnamefont {J}~\bibnamefont {Martin}}, \ and\
  \bibinfo {author} {\bibfnamefont {A.}~\bibnamefont {Yacoby}},\ }\bibfield
  {title} {\enquote {\bibinfo {title} {{Supplementary Information for
  Gate-defined quantum confinement in suspended bilayer graphene}},}\ }\href
  {\doibase 10.1038/ncomms1945} {\bibfield  {journal} {\bibinfo  {journal}
  {Nat. Commun.}\ }\textbf {\bibinfo {volume} {3}},\ \bibinfo {pages} {934}
  (\bibinfo {year} {2012})}\BibitemShut {NoStop}%
\bibitem [{\citenamefont {Eich}\ \emph {et~al.}(2018)\citenamefont {Eich},
  \citenamefont {Pisoni}, \citenamefont {Overweg}, \citenamefont {Kurzmann},
  \citenamefont {Lee}, \citenamefont {Rickhaus}, \citenamefont {Ihn},
  \citenamefont {Ensslin}, \citenamefont {Herman}, \citenamefont {Sigrist},
  \citenamefont {Watanabe},\ and\ \citenamefont {Taniguchi}}]{Eich2018}%
  \BibitemOpen
  \bibfield  {author} {\bibinfo {author} {\bibfnamefont {Marius}\ \bibnamefont
  {Eich}}, \bibinfo {author} {\bibfnamefont {Riccardo}\ \bibnamefont {Pisoni}},
  \bibinfo {author} {\bibfnamefont {Hiske}\ \bibnamefont {Overweg}}, \bibinfo
  {author} {\bibfnamefont {Annika}\ \bibnamefont {Kurzmann}}, \bibinfo {author}
  {\bibfnamefont {Yongjin}\ \bibnamefont {Lee}}, \bibinfo {author}
  {\bibfnamefont {Peter}\ \bibnamefont {Rickhaus}}, \bibinfo {author}
  {\bibfnamefont {Thomas}\ \bibnamefont {Ihn}}, \bibinfo {author}
  {\bibfnamefont {Klaus}\ \bibnamefont {Ensslin}}, \bibinfo {author}
  {\bibfnamefont {Franti{\v{s}}ek}\ \bibnamefont {Herman}}, \bibinfo {author}
  {\bibfnamefont {Manfred}\ \bibnamefont {Sigrist}}, \bibinfo {author}
  {\bibfnamefont {Kenji}\ \bibnamefont {Watanabe}}, \ and\ \bibinfo {author}
  {\bibfnamefont {Takashi}\ \bibnamefont {Taniguchi}},\ }\bibfield  {title}
  {\enquote {\bibinfo {title} {{Spin and Valley States in Gate-Defined Bilayer
  Graphene Quantum Dots}},}\ }\href {\doibase 10.1103/PhysRevX.8.031023}
  {\bibfield  {journal} {\bibinfo  {journal} {Phys. Rev. X}\ }\textbf {\bibinfo
  {volume} {8}},\ \bibinfo {pages} {031023} (\bibinfo {year} {2018})},\ \Eprint
  {http://arxiv.org/abs/1803.02923} {arXiv:1803.02923} \BibitemShut {NoStop}%
\bibitem [{\citenamefont {Choi}\ and\ \citenamefont {Choi}(2019)}]{Choi2019}%
  \BibitemOpen
  \bibfield  {author} {\bibinfo {author} {\bibfnamefont {Young~Woo}\
  \bibnamefont {Choi}}\ and\ \bibinfo {author} {\bibfnamefont {Hyoung~Joon}\
  \bibnamefont {Choi}},\ }\bibfield  {title} {\enquote {\bibinfo {title}
  {{Intrinsic Band Gap and Electrically Tunable Flat Bands in Twisted Double
  Bilayer Graphene}},}\ }\href {http://arxiv.org/abs/1903.00852} {\bibfield
  {journal} {\bibinfo  {journal} {arXiv:1903.00852}\ }\textbf {\bibinfo
  {volume} {1}},\ \bibinfo {pages} {1--7} (\bibinfo {year} {2019})},\ \Eprint
  {http://arxiv.org/abs/1903.00852} {arXiv:1903.00852} \BibitemShut {NoStop}%
\bibitem [{\citenamefont {Chebrolu}\ \emph {et~al.}(2019)\citenamefont
  {Chebrolu}, \citenamefont {Chittari},\ and\ \citenamefont
  {Jung}}]{PhysRevB.99.235417}%
  \BibitemOpen
  \bibfield  {author} {\bibinfo {author} {\bibfnamefont {Narasimha~Raju}\
  \bibnamefont {Chebrolu}}, \bibinfo {author} {\bibfnamefont {Bheema~Lingam}\
  \bibnamefont {Chittari}}, \ and\ \bibinfo {author} {\bibfnamefont {Jeil}\
  \bibnamefont {Jung}},\ }\bibfield  {title} {\enquote {\bibinfo {title} {Flat
  bands in twisted double bilayer graphene},}\ }\href {\doibase
  10.1103/PhysRevB.99.235417} {\bibfield  {journal} {\bibinfo  {journal} {Phys.
  Rev. B}\ }\textbf {\bibinfo {volume} {99}},\ \bibinfo {pages} {235417}
  (\bibinfo {year} {2019})}\BibitemShut {NoStop}%
\bibitem [{\citenamefont {Liu}\ \emph {et~al.}(2019{\natexlab{a}})\citenamefont
  {Liu}, \citenamefont {Ma}, \citenamefont {Gao},\ and\ \citenamefont
  {Dai}}]{PhysRevX.9.031021}%
  \BibitemOpen
  \bibfield  {author} {\bibinfo {author} {\bibfnamefont {Jianpeng}\
  \bibnamefont {Liu}}, \bibinfo {author} {\bibfnamefont {Zhen}\ \bibnamefont
  {Ma}}, \bibinfo {author} {\bibfnamefont {Jinhua}\ \bibnamefont {Gao}}, \ and\
  \bibinfo {author} {\bibfnamefont {Xi}~\bibnamefont {Dai}},\ }\bibfield
  {title} {\enquote {\bibinfo {title} {Quantum valley hall effect, orbital
  magnetism, and anomalous hall effect in twisted multilayer graphene
  systems},}\ }\href {\doibase 10.1103/PhysRevX.9.031021} {\bibfield  {journal}
  {\bibinfo  {journal} {Phys. Rev. X}\ }\textbf {\bibinfo {volume} {9}},\
  \bibinfo {pages} {031021} (\bibinfo {year} {2019}{\natexlab{a}})}\BibitemShut
  {NoStop}%
\bibitem [{\citenamefont {Liu}\ \emph {et~al.}(2019{\natexlab{b}})\citenamefont
  {Liu}, \citenamefont {Hao}, \citenamefont {Khalaf}, \citenamefont {Lee},
  \citenamefont {Watanabe}, \citenamefont {Taniguchi}, \citenamefont
  {Vishwanath},\ and\ \citenamefont {Kim}}]{Liu2019}%
  \BibitemOpen
  \bibfield  {author} {\bibinfo {author} {\bibfnamefont {Xiaomeng}\
  \bibnamefont {Liu}}, \bibinfo {author} {\bibfnamefont {Zeyu}\ \bibnamefont
  {Hao}}, \bibinfo {author} {\bibfnamefont {Eslam}\ \bibnamefont {Khalaf}},
  \bibinfo {author} {\bibfnamefont {Jong~Yeon}\ \bibnamefont {Lee}}, \bibinfo
  {author} {\bibfnamefont {Kenji}\ \bibnamefont {Watanabe}}, \bibinfo {author}
  {\bibfnamefont {Takashi}\ \bibnamefont {Taniguchi}}, \bibinfo {author}
  {\bibfnamefont {Ashvin}\ \bibnamefont {Vishwanath}}, \ and\ \bibinfo {author}
  {\bibfnamefont {Philip}\ \bibnamefont {Kim}},\ }\bibfield  {title} {\enquote
  {\bibinfo {title} {{Spin-polarized Correlated Insulator and Superconductor in
  Twisted Double Bilayer Graphene}},}\ }\href {http://arxiv.org/abs/1903.08130}
  {\bibfield  {journal} {\bibinfo  {journal} {arXiv:1903.08130}\ } (\bibinfo
  {year} {2019}{\natexlab{b}})},\ \Eprint {http://arxiv.org/abs/1903.08130}
  {arXiv:1903.08130} \BibitemShut {NoStop}%
\bibitem [{\citenamefont {Shen}\ \emph {et~al.}(2019)\citenamefont {Shen},
  \citenamefont {Li}, \citenamefont {Wang}, \citenamefont {Zhao}, \citenamefont
  {Tang}, \citenamefont {Liu}, \citenamefont {Tian}, \citenamefont {Chu},
  \citenamefont {Watanabe}, \citenamefont {Taniguchi}, \citenamefont {Yang},
  \citenamefont {Meng}, \citenamefont {Shi},\ and\ \citenamefont
  {Zhang}}]{Shen2019}%
  \BibitemOpen
  \bibfield  {author} {\bibinfo {author} {\bibfnamefont {Cheng}\ \bibnamefont
  {Shen}}, \bibinfo {author} {\bibfnamefont {Na}~\bibnamefont {Li}}, \bibinfo
  {author} {\bibfnamefont {Shuopei}\ \bibnamefont {Wang}}, \bibinfo {author}
  {\bibfnamefont {Yanchong}\ \bibnamefont {Zhao}}, \bibinfo {author}
  {\bibfnamefont {Jian}\ \bibnamefont {Tang}}, \bibinfo {author} {\bibfnamefont
  {Jieying}\ \bibnamefont {Liu}}, \bibinfo {author} {\bibfnamefont {Jinpeng}\
  \bibnamefont {Tian}}, \bibinfo {author} {\bibfnamefont {Yanbang}\
  \bibnamefont {Chu}}, \bibinfo {author} {\bibfnamefont {Kenji}\ \bibnamefont
  {Watanabe}}, \bibinfo {author} {\bibfnamefont {Takashi}\ \bibnamefont
  {Taniguchi}}, \bibinfo {author} {\bibfnamefont {Rong}\ \bibnamefont {Yang}},
  \bibinfo {author} {\bibfnamefont {Zi~Yang}\ \bibnamefont {Meng}}, \bibinfo
  {author} {\bibfnamefont {Dongxia}\ \bibnamefont {Shi}}, \ and\ \bibinfo
  {author} {\bibfnamefont {Guangyu}\ \bibnamefont {Zhang}},\ }\bibfield
  {title} {\enquote {\bibinfo {title} {{Observation of superconductivity with
  Tc onset at 12K in electrically tunable twisted double bilayer graphene}},}\
  }\href {http://arxiv.org/abs/1903.06952} {\bibfield  {journal} {\bibinfo
  {journal} {arXiv:1903.06952}\ } (\bibinfo {year} {2019})},\ \Eprint
  {http://arxiv.org/abs/1903.06952} {arXiv:1903.06952} \BibitemShut {NoStop}%
\bibitem [{\citenamefont {Burg}\ \emph {et~al.}(2019)\citenamefont {Burg},
  \citenamefont {Zhu}, \citenamefont {Taniguchi}, \citenamefont {Watanabe},
  \citenamefont {Macdonald},\ and\ \citenamefont {Tutuc}}]{Burg2019}%
  \BibitemOpen
  \bibfield  {author} {\bibinfo {author} {\bibfnamefont {G~William}\
  \bibnamefont {Burg}}, \bibinfo {author} {\bibfnamefont {Jihang}\ \bibnamefont
  {Zhu}}, \bibinfo {author} {\bibfnamefont {Takashi}\ \bibnamefont
  {Taniguchi}}, \bibinfo {author} {\bibfnamefont {Kenji}\ \bibnamefont
  {Watanabe}}, \bibinfo {author} {\bibfnamefont {Allan~H}\ \bibnamefont
  {Macdonald}}, \ and\ \bibinfo {author} {\bibfnamefont {Emanuel}\ \bibnamefont
  {Tutuc}},\ }\bibfield  {title} {\enquote {\bibinfo {title} {{Correlated
  Insulating States in Twisted Double Bilayer Graphene}},}\ }\href@noop {} {\
  \textbf {\bibinfo {volume} {1}},\ \bibinfo {pages} {1--7} (\bibinfo {year}
  {2019})},\ \Eprint {http://arxiv.org/abs/arXiv:1907.10106v1}
  {arXiv:arXiv:1907.10106v1} \BibitemShut {NoStop}%
\bibitem [{\citenamefont {Kim}\ \emph {et~al.}(2016)\citenamefont {Kim},
  \citenamefont {Yankowitz}, \citenamefont {Fallahazad}, \citenamefont {Kang},
  \citenamefont {Movva}, \citenamefont {Huang}, \citenamefont {Larentis},
  \citenamefont {Corbet}, \citenamefont {Taniguchi}, \citenamefont {Watanabe},
  \citenamefont {Banerjee}, \citenamefont {LeRoy},\ and\ \citenamefont
  {Tutuc}}]{Kim2016}%
  \BibitemOpen
  \bibfield  {author} {\bibinfo {author} {\bibfnamefont {Kyounghwan}\
  \bibnamefont {Kim}}, \bibinfo {author} {\bibfnamefont {Matthew}\ \bibnamefont
  {Yankowitz}}, \bibinfo {author} {\bibfnamefont {Babak}\ \bibnamefont
  {Fallahazad}}, \bibinfo {author} {\bibfnamefont {Sangwoo}\ \bibnamefont
  {Kang}}, \bibinfo {author} {\bibfnamefont {Hema C.~P.}\ \bibnamefont
  {Movva}}, \bibinfo {author} {\bibfnamefont {Shengqiang}\ \bibnamefont
  {Huang}}, \bibinfo {author} {\bibfnamefont {Stefano}\ \bibnamefont
  {Larentis}}, \bibinfo {author} {\bibfnamefont {Chris~M.}\ \bibnamefont
  {Corbet}}, \bibinfo {author} {\bibfnamefont {Takashi}\ \bibnamefont
  {Taniguchi}}, \bibinfo {author} {\bibfnamefont {Kenji}\ \bibnamefont
  {Watanabe}}, \bibinfo {author} {\bibfnamefont {Sanjay~K.}\ \bibnamefont
  {Banerjee}}, \bibinfo {author} {\bibfnamefont {Brian~J.}\ \bibnamefont
  {LeRoy}}, \ and\ \bibinfo {author} {\bibfnamefont {Emanuel}\ \bibnamefont
  {Tutuc}},\ }\bibfield  {title} {\enquote {\bibinfo {title} {{van der Waals
  Heterostructures with High Accuracy Rotational Alignment}},}\ }\href
  {\doibase 10.1021/acs.nanolett.5b05263} {\bibfield  {journal} {\bibinfo
  {journal} {Nano Lett.}\ }\textbf {\bibinfo {volume} {16}},\ \bibinfo {pages}
  {1989--1995} (\bibinfo {year} {2016})}\BibitemShut {NoStop}%
\bibitem [{\citenamefont {Kim}\ \emph {et~al.}(2017)\citenamefont {Kim},
  \citenamefont {DaSilva}, \citenamefont {Huang}, \citenamefont {Fallahazad},
  \citenamefont {Larentis}, \citenamefont {Taniguchi}, \citenamefont
  {Watanabe}, \citenamefont {Leroy}, \citenamefont {Macdonald}, \citenamefont
  {Tutuc}, \citenamefont {Kim},\ and\ \citenamefont {Novoselov}}]{Kim2017}%
  \BibitemOpen
  \bibfield  {author} {\bibinfo {author} {\bibfnamefont {Kyounghwan}\
  \bibnamefont {Kim}}, \bibinfo {author} {\bibfnamefont {Ashley}\ \bibnamefont
  {DaSilva}}, \bibinfo {author} {\bibfnamefont {Shengqiang}\ \bibnamefont
  {Huang}}, \bibinfo {author} {\bibfnamefont {Babak}\ \bibnamefont
  {Fallahazad}}, \bibinfo {author} {\bibfnamefont {Stefano}\ \bibnamefont
  {Larentis}}, \bibinfo {author} {\bibfnamefont {Takashi}\ \bibnamefont
  {Taniguchi}}, \bibinfo {author} {\bibfnamefont {Kenji}\ \bibnamefont
  {Watanabe}}, \bibinfo {author} {\bibfnamefont {Brian~J.}\ \bibnamefont
  {Leroy}}, \bibinfo {author} {\bibfnamefont {Allan~H.}\ \bibnamefont
  {Macdonald}}, \bibinfo {author} {\bibfnamefont {Emanuel}\ \bibnamefont
  {Tutuc}}, \bibinfo {author} {\bibfnamefont {Philip}\ \bibnamefont {Kim}}, \
  and\ \bibinfo {author} {\bibfnamefont {Konstantin~S}\ \bibnamefont
  {Novoselov}},\ }\bibfield  {title} {\enquote {\bibinfo {title} {{Tunable
  moir{\'{e}} bands and strong correlations in small-twist-angle bilayer
  graphene}},}\ }\href {\doibase 10.1073/pnas.1620140114} {\bibfield  {journal}
  {\bibinfo  {journal} {Proc. Natl. Acad. Sci.}\ }\textbf {\bibinfo {volume}
  {114}},\ \bibinfo {pages} {3364--3369} (\bibinfo {year} {2017})}\BibitemShut
  {NoStop}%
\bibitem [{\citenamefont {Wang}\ \emph {et~al.}(2013)\citenamefont {Wang},
  \citenamefont {Meric}, \citenamefont {Huang}, \citenamefont {Gao},
  \citenamefont {Gao}, \citenamefont {Tran}, \citenamefont {Taniguchi},
  \citenamefont {Watanabe}, \citenamefont {Campos}, \citenamefont {Muller},
  \citenamefont {Guo}, \citenamefont {Kim}, \citenamefont {Hone}, \citenamefont
  {Shepard},\ and\ \citenamefont {Dean}}]{Wang2013}%
  \BibitemOpen
  \bibfield  {author} {\bibinfo {author} {\bibfnamefont {L}~\bibnamefont
  {Wang}}, \bibinfo {author} {\bibfnamefont {I}~\bibnamefont {Meric}}, \bibinfo
  {author} {\bibfnamefont {P~Y}\ \bibnamefont {Huang}}, \bibinfo {author}
  {\bibfnamefont {Q}~\bibnamefont {Gao}}, \bibinfo {author} {\bibfnamefont
  {Y}~\bibnamefont {Gao}}, \bibinfo {author} {\bibfnamefont {H}~\bibnamefont
  {Tran}}, \bibinfo {author} {\bibfnamefont {T}~\bibnamefont {Taniguchi}},
  \bibinfo {author} {\bibfnamefont {K}~\bibnamefont {Watanabe}}, \bibinfo
  {author} {\bibfnamefont {L~M}\ \bibnamefont {Campos}}, \bibinfo {author}
  {\bibfnamefont {D~A}\ \bibnamefont {Muller}}, \bibinfo {author}
  {\bibfnamefont {J}~\bibnamefont {Guo}}, \bibinfo {author} {\bibfnamefont
  {P}~\bibnamefont {Kim}}, \bibinfo {author} {\bibfnamefont {J}~\bibnamefont
  {Hone}}, \bibinfo {author} {\bibfnamefont {K~L}\ \bibnamefont {Shepard}}, \
  and\ \bibinfo {author} {\bibfnamefont {C~R}\ \bibnamefont {Dean}},\
  }\bibfield  {title} {\enquote {\bibinfo {title} {{One-dimensional electrical
  contact to a two-dimensional material.}}}\ }\href {\doibase
  10.1126/science.1244358} {\bibfield  {journal} {\bibinfo  {journal}
  {Science}\ }\textbf {\bibinfo {volume} {342}},\ \bibinfo {pages} {614--7}
  (\bibinfo {year} {2013})}\BibitemShut {NoStop}%
\bibitem [{\citenamefont {Zibrov}\ \emph {et~al.}(2017)\citenamefont {Zibrov},
  \citenamefont {Kometter}, \citenamefont {Zhou}, \citenamefont {Spanton},
  \citenamefont {Taniguchi}, \citenamefont {Watanabe}, \citenamefont
  {Zaletel},\ and\ \citenamefont {Young}}]{Zibrov2017a}%
  \BibitemOpen
  \bibfield  {author} {\bibinfo {author} {\bibfnamefont {A.~A.}\ \bibnamefont
  {Zibrov}}, \bibinfo {author} {\bibfnamefont {C.}~\bibnamefont {Kometter}},
  \bibinfo {author} {\bibfnamefont {H.}~\bibnamefont {Zhou}}, \bibinfo {author}
  {\bibfnamefont {E.~M.}\ \bibnamefont {Spanton}}, \bibinfo {author}
  {\bibfnamefont {T.}~\bibnamefont {Taniguchi}}, \bibinfo {author}
  {\bibfnamefont {K.}~\bibnamefont {Watanabe}}, \bibinfo {author}
  {\bibfnamefont {M.~P.}\ \bibnamefont {Zaletel}}, \ and\ \bibinfo {author}
  {\bibfnamefont {A.~F.}\ \bibnamefont {Young}},\ }\bibfield  {title} {\enquote
  {\bibinfo {title} {{Tunable interacting composite fermion phases in a
  half-filled bilayer-graphene Landau level}},}\ }\href {\doibase
  10.1038/nature23893} {\bibfield  {journal} {\bibinfo  {journal} {Nature}\
  }\textbf {\bibinfo {volume} {549}},\ \bibinfo {pages} {360--364} (\bibinfo
  {year} {2017})}\BibitemShut {NoStop}%
\bibitem [{\citenamefont {Li}\ \emph {et~al.}(2016)\citenamefont {Li},
  \citenamefont {Tan}, \citenamefont {Zou}, \citenamefont {Stabile},
  \citenamefont {Seiwell}, \citenamefont {Watanabe}, \citenamefont {Taniguchi},
  \citenamefont {Louie},\ and\ \citenamefont {Zhu}}]{Li2016}%
  \BibitemOpen
  \bibfield  {author} {\bibinfo {author} {\bibfnamefont {J.}~\bibnamefont
  {Li}}, \bibinfo {author} {\bibfnamefont {L.~Z.}\ \bibnamefont {Tan}},
  \bibinfo {author} {\bibfnamefont {K.}~\bibnamefont {Zou}}, \bibinfo {author}
  {\bibfnamefont {A.~A.}\ \bibnamefont {Stabile}}, \bibinfo {author}
  {\bibfnamefont {D.~J.}\ \bibnamefont {Seiwell}}, \bibinfo {author}
  {\bibfnamefont {K.}~\bibnamefont {Watanabe}}, \bibinfo {author}
  {\bibfnamefont {T.}~\bibnamefont {Taniguchi}}, \bibinfo {author}
  {\bibfnamefont {Steven~G.}\ \bibnamefont {Louie}}, \ and\ \bibinfo {author}
  {\bibfnamefont {J.}~\bibnamefont {Zhu}},\ }\bibfield  {title} {\enquote
  {\bibinfo {title} {Effective mass in bilayer graphene at low carrier
  densities: The role of potential disorder and electron-electron
  interaction},}\ }\href {\doibase 10.1103/PhysRevB.94.161406} {\bibfield
  {journal} {\bibinfo  {journal} {Phys. Rev. B}\ }\textbf {\bibinfo {volume}
  {94}},\ \bibinfo {pages} {161406} (\bibinfo {year} {2016})}\BibitemShut
  {NoStop}%
\bibitem [{\citenamefont {Young}\ and\ \citenamefont {Kim}(2009)}]{Young2009}%
  \BibitemOpen
  \bibfield  {author} {\bibinfo {author} {\bibfnamefont {Andrea~F}\
  \bibnamefont {Young}}\ and\ \bibinfo {author} {\bibfnamefont {Philip}\
  \bibnamefont {Kim}},\ }\bibfield  {title} {\enquote {\bibinfo {title}
  {{Quantum interference and Klein tunnelling in graphene heterojunctions}},}\
  }\href {http://dx.doi.org/10.1038/nphys1198 http://10.0.4.14/nphys1198
  https://www.nature.com/articles/nphys1198{\#}supplementary-information}
  {\bibfield  {journal} {\bibinfo  {journal} {Nat. Phys.}\ }\textbf {\bibinfo
  {volume} {5}},\ \bibinfo {pages} {222} (\bibinfo {year} {2009})}\BibitemShut
  {NoStop}%
\bibitem [{\citenamefont {Rickhaus}\ \emph {et~al.}(2013)\citenamefont
  {Rickhaus}, \citenamefont {Maurand}, \citenamefont {Liu}, \citenamefont
  {Weiss}, \citenamefont {Richter},\ and\ \citenamefont
  {Sch{\"{o}}nenberger}}]{Rickhaus2013}%
  \BibitemOpen
  \bibfield  {author} {\bibinfo {author} {\bibfnamefont {Peter}\ \bibnamefont
  {Rickhaus}}, \bibinfo {author} {\bibfnamefont {Romain}\ \bibnamefont
  {Maurand}}, \bibinfo {author} {\bibfnamefont {Ming-Hao}\ \bibnamefont {Liu}},
  \bibinfo {author} {\bibfnamefont {Markus}\ \bibnamefont {Weiss}}, \bibinfo
  {author} {\bibfnamefont {Klaus}\ \bibnamefont {Richter}}, \ and\ \bibinfo
  {author} {\bibfnamefont {Christian}\ \bibnamefont {Sch{\"{o}}nenberger}},\
  }\bibfield  {title} {\enquote {\bibinfo {title} {{Ballistic interferences in
  suspended graphene}},}\ }\href {http://dx.doi.org/10.1038/ncomms3342}
  {\bibfield  {journal} {\bibinfo  {journal} {Nat. Commun.}\ }\textbf {\bibinfo
  {volume} {4}},\ \bibinfo {pages} {3342} (\bibinfo {year} {2013})}\BibitemShut
  {NoStop}%
\bibitem [{\citenamefont {Varlet}\ \emph {et~al.}(2014)\citenamefont {Varlet},
  \citenamefont {Liu}, \citenamefont {Krueckl}, \citenamefont {Bischoff},
  \citenamefont {Simonet}, \citenamefont {Watanabe}, \citenamefont {Taniguchi},
  \citenamefont {Richter}, \citenamefont {Ensslin},\ and\ \citenamefont
  {Ihn}}]{Varlet2014}%
  \BibitemOpen
  \bibfield  {author} {\bibinfo {author} {\bibfnamefont {Anastasia}\
  \bibnamefont {Varlet}}, \bibinfo {author} {\bibfnamefont {Ming-Hao}\
  \bibnamefont {Liu}}, \bibinfo {author} {\bibfnamefont {Viktor}\ \bibnamefont
  {Krueckl}}, \bibinfo {author} {\bibfnamefont {Dominik}\ \bibnamefont
  {Bischoff}}, \bibinfo {author} {\bibfnamefont {Pauline}\ \bibnamefont
  {Simonet}}, \bibinfo {author} {\bibfnamefont {Kenji}\ \bibnamefont
  {Watanabe}}, \bibinfo {author} {\bibfnamefont {Takashi}\ \bibnamefont
  {Taniguchi}}, \bibinfo {author} {\bibfnamefont {Klaus}\ \bibnamefont
  {Richter}}, \bibinfo {author} {\bibfnamefont {Klaus}\ \bibnamefont
  {Ensslin}}, \ and\ \bibinfo {author} {\bibfnamefont {Thomas}\ \bibnamefont
  {Ihn}},\ }\bibfield  {title} {\enquote {\bibinfo {title} {{Fabry-P{\'{e}}rot
  Interference in Gapped Bilayer Graphene with Broken Anti-Klein Tunneling}},}\
  }\href {\doibase 10.1103/PhysRevLett.113.116601} {\bibfield  {journal}
  {\bibinfo  {journal} {Phys. Rev. Lett.}\ }\textbf {\bibinfo {volume} {113}},\
  \bibinfo {pages} {116601} (\bibinfo {year} {2014})}\BibitemShut {NoStop}%
\bibitem [{\citenamefont {Handschin}\ \emph {et~al.}(2017)\citenamefont
  {Handschin}, \citenamefont {Makk}, \citenamefont {Rickhaus}, \citenamefont
  {Maurand}, \citenamefont {Watanabe}, \citenamefont {Taniguchi}, \citenamefont
  {Richter}, \citenamefont {Liu},\ and\ \citenamefont
  {Sch{\"{o}}nenberger}}]{Handschin2017}%
  \BibitemOpen
  \bibfield  {author} {\bibinfo {author} {\bibfnamefont {Clevin}\ \bibnamefont
  {Handschin}}, \bibinfo {author} {\bibfnamefont {P{\'{e}}ter}\ \bibnamefont
  {Makk}}, \bibinfo {author} {\bibfnamefont {Peter}\ \bibnamefont {Rickhaus}},
  \bibinfo {author} {\bibfnamefont {Romain}\ \bibnamefont {Maurand}}, \bibinfo
  {author} {\bibfnamefont {Kenji}\ \bibnamefont {Watanabe}}, \bibinfo {author}
  {\bibfnamefont {Takashi}\ \bibnamefont {Taniguchi}}, \bibinfo {author}
  {\bibfnamefont {Klaus}\ \bibnamefont {Richter}}, \bibinfo {author}
  {\bibfnamefont {Ming-Hao}\ \bibnamefont {Liu}}, \ and\ \bibinfo {author}
  {\bibfnamefont {Christian}\ \bibnamefont {Sch{\"{o}}nenberger}},\ }\bibfield
  {title} {\enquote {\bibinfo {title} {{Giant Valley-Isospin Conductance
  Oscillations in Ballistic Graphene}},}\ }\href {\doibase
  10.1021/acs.nanolett.7b01964} {\bibfield  {journal} {\bibinfo  {journal}
  {Nano Lett.}\ }\textbf {\bibinfo {volume} {17}},\ \bibinfo {pages}
  {5389--5393} (\bibinfo {year} {2017})}\BibitemShut {NoStop}%
\bibitem [{\citenamefont {Toy}\ \emph {et~al.}(1977)\citenamefont {Toy},
  \citenamefont {Dresselhaus},\ and\ \citenamefont
  {Dresselhaus}}]{PhysRevB.15.4077}%
  \BibitemOpen
  \bibfield  {author} {\bibinfo {author} {\bibfnamefont {W.~W.}\ \bibnamefont
  {Toy}}, \bibinfo {author} {\bibfnamefont {M.~S.}\ \bibnamefont
  {Dresselhaus}}, \ and\ \bibinfo {author} {\bibfnamefont {G.}~\bibnamefont
  {Dresselhaus}},\ }\bibfield  {title} {\enquote {\bibinfo {title} {Minority
  carriers in graphite and the $h$-point magnetoreflection spectra},}\ }\href
  {\doibase 10.1103/PhysRevB.15.4077} {\bibfield  {journal} {\bibinfo
  {journal} {Phys. Rev. B}\ }\textbf {\bibinfo {volume} {15}},\ \bibinfo
  {pages} {4077--4090} (\bibinfo {year} {1977})}\BibitemShut {NoStop}%
\bibitem [{\citenamefont {Gr\"uneis}\ \emph {et~al.}(2008)\citenamefont
  {Gr\"uneis}, \citenamefont {Attaccalite}, \citenamefont {Wirtz},
  \citenamefont {Shiozawa}, \citenamefont {Saito}, \citenamefont {Pichler},\
  and\ \citenamefont {Rubio}}]{PhysRevB.78.205425}%
  \BibitemOpen
  \bibfield  {author} {\bibinfo {author} {\bibfnamefont {A.}~\bibnamefont
  {Gr\"uneis}}, \bibinfo {author} {\bibfnamefont {C.}~\bibnamefont
  {Attaccalite}}, \bibinfo {author} {\bibfnamefont {L.}~\bibnamefont {Wirtz}},
  \bibinfo {author} {\bibfnamefont {H.}~\bibnamefont {Shiozawa}}, \bibinfo
  {author} {\bibfnamefont {R.}~\bibnamefont {Saito}}, \bibinfo {author}
  {\bibfnamefont {T.}~\bibnamefont {Pichler}}, \ and\ \bibinfo {author}
  {\bibfnamefont {A.}~\bibnamefont {Rubio}},\ }\bibfield  {title} {\enquote
  {\bibinfo {title} {Tight-binding description of the quasiparticle dispersion
  of graphite and few-layer graphene},}\ }\href {\doibase
  10.1103/PhysRevB.78.205425} {\bibfield  {journal} {\bibinfo  {journal} {Phys.
  Rev. B}\ }\textbf {\bibinfo {volume} {78}},\ \bibinfo {pages} {205425}
  (\bibinfo {year} {2008})}\BibitemShut {NoStop}%
\bibitem [{\citenamefont {Guinea}\ and\ \citenamefont
  {Walet}(2018)}]{Guinea2018}%
  \BibitemOpen
  \bibfield  {author} {\bibinfo {author} {\bibfnamefont {Francisco}\
  \bibnamefont {Guinea}}\ and\ \bibinfo {author} {\bibfnamefont {Niels~R.}\
  \bibnamefont {Walet}},\ }\bibfield  {title} {\enquote {\bibinfo {title}
  {Electrostatic effects, band distortions, and superconductivity in twisted
  graphene bilayers},}\ }\href {\doibase 10.1073/pnas.1810947115} {\bibfield
  {journal} {\bibinfo  {journal} {Proceedings of the National Academy of
  Sciences}\ }\textbf {\bibinfo {volume} {115}},\ \bibinfo {pages}
  {13174--13179} (\bibinfo {year} {2018})}\BibitemShut {NoStop}%
\bibitem [{\citenamefont {Cysne}\ \emph {et~al.}(2018)\citenamefont {Cysne},
  \citenamefont {Ferreira},\ and\ \citenamefont
  {Rappoport}}]{PhysRevB.98.045407}%
  \BibitemOpen
  \bibfield  {author} {\bibinfo {author} {\bibfnamefont {Tarik~P.}\
  \bibnamefont {Cysne}}, \bibinfo {author} {\bibfnamefont {Aires}\ \bibnamefont
  {Ferreira}}, \ and\ \bibinfo {author} {\bibfnamefont {Tatiana~G.}\
  \bibnamefont {Rappoport}},\ }\bibfield  {title} {\enquote {\bibinfo {title}
  {Crystal-field effects in graphene with interface-induced spin-orbit
  coupling},}\ }\href {\doibase 10.1103/PhysRevB.98.045407} {\bibfield
  {journal} {\bibinfo  {journal} {Phys. Rev. B}\ }\textbf {\bibinfo {volume}
  {98}},\ \bibinfo {pages} {045407} (\bibinfo {year} {2018})}\BibitemShut
  {NoStop}%
\bibitem [{\citenamefont {{Choi}}\ and\ \citenamefont
  {{Choi}}(2019)}]{2019arXiv190300852C}%
  \BibitemOpen
  \bibfield  {author} {\bibinfo {author} {\bibfnamefont {Young~Woo}\
  \bibnamefont {{Choi}}}\ and\ \bibinfo {author} {\bibfnamefont {Hyoung~Joon}\
  \bibnamefont {{Choi}}},\ }\bibfield  {title} {\enquote {\bibinfo {title}
  {{Intrinsic Band Gap and Electrically Tunable Flat Bands in Twisted Double
  Bilayer Graphene}},}\ }\href@noop {} {\bibfield  {journal} {\bibinfo
  {journal} {arXiv e-prints}\ ,\ \bibinfo {eid} {arXiv:1903.00852}} (\bibinfo
  {year} {2019})},\ \Eprint {http://arxiv.org/abs/1903.00852} {arXiv:1903.00852
  [cond-mat.mes-hall]} \BibitemShut {NoStop}%
\bibitem [{\citenamefont {Koshino}(2019)}]{PhysRevB.99.235406}%
  \BibitemOpen
  \bibfield  {author} {\bibinfo {author} {\bibfnamefont {Mikito}\ \bibnamefont
  {Koshino}},\ }\bibfield  {title} {\enquote {\bibinfo {title} {Band structure
  and topological properties of twisted double bilayer graphene},}\ }\href
  {\doibase 10.1103/PhysRevB.99.235406} {\bibfield  {journal} {\bibinfo
  {journal} {Phys. Rev. B}\ }\textbf {\bibinfo {volume} {99}},\ \bibinfo
  {pages} {235406} (\bibinfo {year} {2019})}\BibitemShut {NoStop}%
\bibitem [{\citenamefont {Grimme}(2006)}]{Grimme2006}%
  \BibitemOpen
  \bibfield  {author} {\bibinfo {author} {\bibfnamefont {Stefan}\ \bibnamefont
  {Grimme}},\ }\bibfield  {title} {\enquote {\bibinfo {title} {Semiempirical
  {GGA}-type density functional constructed with a long-range dispersion
  correction},}\ }\href {\doibase 10.1002/jcc.20495} {\bibfield  {journal}
  {\bibinfo  {journal} {Journal of Computational Chemistry}\ }\textbf {\bibinfo
  {volume} {27}},\ \bibinfo {pages} {1787--1799} (\bibinfo {year}
  {2006})}\BibitemShut {NoStop}%
\bibitem [{\citenamefont {Barone}\ \emph {et~al.}(2009)\citenamefont {Barone},
  \citenamefont {Casarin}, \citenamefont {Forrer}, \citenamefont {Pavone},
  \citenamefont {Sambi},\ and\ \citenamefont {Vittadini}}]{Barone2009}%
  \BibitemOpen
  \bibfield  {author} {\bibinfo {author} {\bibfnamefont {Vincenzo}\
  \bibnamefont {Barone}}, \bibinfo {author} {\bibfnamefont {Maurizio}\
  \bibnamefont {Casarin}}, \bibinfo {author} {\bibfnamefont {Daniel}\
  \bibnamefont {Forrer}}, \bibinfo {author} {\bibfnamefont {Michele}\
  \bibnamefont {Pavone}}, \bibinfo {author} {\bibfnamefont {Mauro}\
  \bibnamefont {Sambi}}, \ and\ \bibinfo {author} {\bibfnamefont {Andrea}\
  \bibnamefont {Vittadini}},\ }\bibfield  {title} {\enquote {\bibinfo {title}
  {Role and effective treatment of dispersive forces in materials: Polyethylene
  and graphite crystals as test cases},}\ }\href {\doibase 10.1002/jcc.21112}
  {\bibfield  {journal} {\bibinfo  {journal} {Journal of Computational
  Chemistry}\ }\textbf {\bibinfo {volume} {30}},\ \bibinfo {pages} {934--939}
  (\bibinfo {year} {2009})}\BibitemShut {NoStop}%
\bibitem [{\citenamefont {Giannozzi}\ \emph {et~al.}(2009)\citenamefont
  {Giannozzi}, \citenamefont {Baroni}, \citenamefont {Bonini}, \citenamefont
  {Calandra}, \citenamefont {Car}, \citenamefont {Cavazzoni}, \citenamefont
  {Ceresoli}, \citenamefont {Chiarotti}, \citenamefont {Cococcioni},
  \citenamefont {Dabo}, \citenamefont {Corso}, \citenamefont {de~Gironcoli},
  \citenamefont {Fabris}, \citenamefont {Fratesi}, \citenamefont {Gebauer},
  \citenamefont {Gerstmann}, \citenamefont {Gougoussis}, \citenamefont
  {Kokalj}, \citenamefont {Lazzeri}, \citenamefont {Martin-Samos},
  \citenamefont {Marzari}, \citenamefont {Mauri}, \citenamefont {Mazzarello},
  \citenamefont {Paolini}, \citenamefont {Pasquarello}, \citenamefont
  {Paulatto}, \citenamefont {Sbraccia}, \citenamefont {Scandolo}, \citenamefont
  {Sclauzero}, \citenamefont {Seitsonen}, \citenamefont {Smogunov},
  \citenamefont {Umari},\ and\ \citenamefont {Wentzcovitch}}]{Giannozzi2009}%
  \BibitemOpen
  \bibfield  {author} {\bibinfo {author} {\bibfnamefont {Paolo}\ \bibnamefont
  {Giannozzi}}, \bibinfo {author} {\bibfnamefont {Stefano}\ \bibnamefont
  {Baroni}}, \bibinfo {author} {\bibfnamefont {Nicola}\ \bibnamefont {Bonini}},
  \bibinfo {author} {\bibfnamefont {Matteo}\ \bibnamefont {Calandra}}, \bibinfo
  {author} {\bibfnamefont {Roberto}\ \bibnamefont {Car}}, \bibinfo {author}
  {\bibfnamefont {Carlo}\ \bibnamefont {Cavazzoni}}, \bibinfo {author}
  {\bibfnamefont {Davide}\ \bibnamefont {Ceresoli}}, \bibinfo {author}
  {\bibfnamefont {Guido~L}\ \bibnamefont {Chiarotti}}, \bibinfo {author}
  {\bibfnamefont {Matteo}\ \bibnamefont {Cococcioni}}, \bibinfo {author}
  {\bibfnamefont {Ismaila}\ \bibnamefont {Dabo}}, \bibinfo {author}
  {\bibfnamefont {Andrea~Dal}\ \bibnamefont {Corso}}, \bibinfo {author}
  {\bibfnamefont {Stefano}\ \bibnamefont {de~Gironcoli}}, \bibinfo {author}
  {\bibfnamefont {Stefano}\ \bibnamefont {Fabris}}, \bibinfo {author}
  {\bibfnamefont {Guido}\ \bibnamefont {Fratesi}}, \bibinfo {author}
  {\bibfnamefont {Ralph}\ \bibnamefont {Gebauer}}, \bibinfo {author}
  {\bibfnamefont {Uwe}\ \bibnamefont {Gerstmann}}, \bibinfo {author}
  {\bibfnamefont {Christos}\ \bibnamefont {Gougoussis}}, \bibinfo {author}
  {\bibfnamefont {Anton}\ \bibnamefont {Kokalj}}, \bibinfo {author}
  {\bibfnamefont {Michele}\ \bibnamefont {Lazzeri}}, \bibinfo {author}
  {\bibfnamefont {Layla}\ \bibnamefont {Martin-Samos}}, \bibinfo {author}
  {\bibfnamefont {Nicola}\ \bibnamefont {Marzari}}, \bibinfo {author}
  {\bibfnamefont {Francesco}\ \bibnamefont {Mauri}}, \bibinfo {author}
  {\bibfnamefont {Riccardo}\ \bibnamefont {Mazzarello}}, \bibinfo {author}
  {\bibfnamefont {Stefano}\ \bibnamefont {Paolini}}, \bibinfo {author}
  {\bibfnamefont {Alfredo}\ \bibnamefont {Pasquarello}}, \bibinfo {author}
  {\bibfnamefont {Lorenzo}\ \bibnamefont {Paulatto}}, \bibinfo {author}
  {\bibfnamefont {Carlo}\ \bibnamefont {Sbraccia}}, \bibinfo {author}
  {\bibfnamefont {Sandro}\ \bibnamefont {Scandolo}}, \bibinfo {author}
  {\bibfnamefont {Gabriele}\ \bibnamefont {Sclauzero}}, \bibinfo {author}
  {\bibfnamefont {Ari~P}\ \bibnamefont {Seitsonen}}, \bibinfo {author}
  {\bibfnamefont {Alexander}\ \bibnamefont {Smogunov}}, \bibinfo {author}
  {\bibfnamefont {Paolo}\ \bibnamefont {Umari}}, \ and\ \bibinfo {author}
  {\bibfnamefont {Renata~M}\ \bibnamefont {Wentzcovitch}},\ }\bibfield  {title}
  {\enquote {\bibinfo {title} {{QUANTUM} {ESPRESSO}: a modular and open-source
  software project for quantum simulations of materials},}\ }\href {\doibase
  10.1088/0953-8984/21/39/395502} {\bibfield  {journal} {\bibinfo  {journal}
  {Journal of Physics: Condensed Matter}\ }\textbf {\bibinfo {volume} {21}},\
  \bibinfo {pages} {395502} (\bibinfo {year} {2009})}\BibitemShut {NoStop}%
\bibitem [{\citenamefont {Giannozzi}\ \emph {et~al.}(2017)\citenamefont
  {Giannozzi}, \citenamefont {Andreussi}, \citenamefont {Brumme}, \citenamefont
  {Bunau}, \citenamefont {Nardelli}, \citenamefont {Calandra}, \citenamefont
  {Car}, \citenamefont {Cavazzoni}, \citenamefont {Ceresoli}, \citenamefont
  {Cococcioni}, \citenamefont {Colonna}, \citenamefont {Carnimeo},
  \citenamefont {Corso}, \citenamefont {de~Gironcoli}, \citenamefont {Delugas},
  \citenamefont {DiStasio}, \citenamefont {Ferretti}, \citenamefont {Floris},
  \citenamefont {Fratesi}, \citenamefont {Fugallo}, \citenamefont {Gebauer},
  \citenamefont {Gerstmann}, \citenamefont {Giustino}, \citenamefont {Gorni},
  \citenamefont {Jia}, \citenamefont {Kawamura}, \citenamefont {Ko},
  \citenamefont {Kokalj}, \citenamefont {K\"{u}{\c{c}}\"{u}kbenli},
  \citenamefont {Lazzeri}, \citenamefont {Marsili}, \citenamefont {Marzari},
  \citenamefont {Mauri}, \citenamefont {Nguyen}, \citenamefont {Nguyen},
  \citenamefont {de-la Roza}, \citenamefont {Paulatto}, \citenamefont
  {Ponc{\'{e}}}, \citenamefont {Rocca}, \citenamefont {Sabatini}, \citenamefont
  {Santra}, \citenamefont {Schlipf}, \citenamefont {Seitsonen}, \citenamefont
  {Smogunov}, \citenamefont {Timrov}, \citenamefont {Thonhauser}, \citenamefont
  {Umari}, \citenamefont {Vast}, \citenamefont {Wu},\ and\ \citenamefont
  {Baroni}}]{Giannozzi2017}%
  \BibitemOpen
  \bibfield  {author} {\bibinfo {author} {\bibfnamefont {P}~\bibnamefont
  {Giannozzi}}, \bibinfo {author} {\bibfnamefont {O}~\bibnamefont {Andreussi}},
  \bibinfo {author} {\bibfnamefont {T}~\bibnamefont {Brumme}}, \bibinfo
  {author} {\bibfnamefont {O}~\bibnamefont {Bunau}}, \bibinfo {author}
  {\bibfnamefont {M~Buongiorno}\ \bibnamefont {Nardelli}}, \bibinfo {author}
  {\bibfnamefont {M}~\bibnamefont {Calandra}}, \bibinfo {author} {\bibfnamefont
  {R}~\bibnamefont {Car}}, \bibinfo {author} {\bibfnamefont {C}~\bibnamefont
  {Cavazzoni}}, \bibinfo {author} {\bibfnamefont {D}~\bibnamefont {Ceresoli}},
  \bibinfo {author} {\bibfnamefont {M}~\bibnamefont {Cococcioni}}, \bibinfo
  {author} {\bibfnamefont {N}~\bibnamefont {Colonna}}, \bibinfo {author}
  {\bibfnamefont {I}~\bibnamefont {Carnimeo}}, \bibinfo {author} {\bibfnamefont
  {A~Dal}\ \bibnamefont {Corso}}, \bibinfo {author} {\bibfnamefont
  {S}~\bibnamefont {de~Gironcoli}}, \bibinfo {author} {\bibfnamefont
  {P}~\bibnamefont {Delugas}}, \bibinfo {author} {\bibfnamefont {R~A}\
  \bibnamefont {DiStasio}}, \bibinfo {author} {\bibfnamefont {A}~\bibnamefont
  {Ferretti}}, \bibinfo {author} {\bibfnamefont {A}~\bibnamefont {Floris}},
  \bibinfo {author} {\bibfnamefont {G}~\bibnamefont {Fratesi}}, \bibinfo
  {author} {\bibfnamefont {G}~\bibnamefont {Fugallo}}, \bibinfo {author}
  {\bibfnamefont {R}~\bibnamefont {Gebauer}}, \bibinfo {author} {\bibfnamefont
  {U}~\bibnamefont {Gerstmann}}, \bibinfo {author} {\bibfnamefont
  {F}~\bibnamefont {Giustino}}, \bibinfo {author} {\bibfnamefont
  {T}~\bibnamefont {Gorni}}, \bibinfo {author} {\bibfnamefont {J}~\bibnamefont
  {Jia}}, \bibinfo {author} {\bibfnamefont {M}~\bibnamefont {Kawamura}},
  \bibinfo {author} {\bibfnamefont {H-Y}\ \bibnamefont {Ko}}, \bibinfo {author}
  {\bibfnamefont {A}~\bibnamefont {Kokalj}}, \bibinfo {author} {\bibfnamefont
  {E}~\bibnamefont {K\"{u}{\c{c}}\"{u}kbenli}}, \bibinfo {author}
  {\bibfnamefont {M}~\bibnamefont {Lazzeri}}, \bibinfo {author} {\bibfnamefont
  {M}~\bibnamefont {Marsili}}, \bibinfo {author} {\bibfnamefont
  {N}~\bibnamefont {Marzari}}, \bibinfo {author} {\bibfnamefont
  {F}~\bibnamefont {Mauri}}, \bibinfo {author} {\bibfnamefont {N~L}\
  \bibnamefont {Nguyen}}, \bibinfo {author} {\bibfnamefont {H-V}\ \bibnamefont
  {Nguyen}}, \bibinfo {author} {\bibfnamefont {A~Otero}\ \bibnamefont {de-la
  Roza}}, \bibinfo {author} {\bibfnamefont {L}~\bibnamefont {Paulatto}},
  \bibinfo {author} {\bibfnamefont {S}~\bibnamefont {Ponc{\'{e}}}}, \bibinfo
  {author} {\bibfnamefont {D}~\bibnamefont {Rocca}}, \bibinfo {author}
  {\bibfnamefont {R}~\bibnamefont {Sabatini}}, \bibinfo {author} {\bibfnamefont
  {B}~\bibnamefont {Santra}}, \bibinfo {author} {\bibfnamefont {M}~\bibnamefont
  {Schlipf}}, \bibinfo {author} {\bibfnamefont {A~P}\ \bibnamefont
  {Seitsonen}}, \bibinfo {author} {\bibfnamefont {A}~\bibnamefont {Smogunov}},
  \bibinfo {author} {\bibfnamefont {I}~\bibnamefont {Timrov}}, \bibinfo
  {author} {\bibfnamefont {T}~\bibnamefont {Thonhauser}}, \bibinfo {author}
  {\bibfnamefont {P}~\bibnamefont {Umari}}, \bibinfo {author} {\bibfnamefont
  {N}~\bibnamefont {Vast}}, \bibinfo {author} {\bibfnamefont {X}~\bibnamefont
  {Wu}}, \ and\ \bibinfo {author} {\bibfnamefont {S}~\bibnamefont {Baroni}},\
  }\bibfield  {title} {\enquote {\bibinfo {title} {Advanced capabilities for
  materials modelling with quantum {ESPRESSO}},}\ }\href {\doibase
  10.1088/1361-648x/aa8f79} {\bibfield  {journal} {\bibinfo  {journal} {Journal
  of Physics: Condensed Matter}\ }\textbf {\bibinfo {volume} {29}},\ \bibinfo
  {pages} {465901} (\bibinfo {year} {2017})}\BibitemShut {NoStop}%
\bibitem [{\citenamefont {Vanderbilt}(1990)}]{PhysRevB.41.7892}%
  \BibitemOpen
  \bibfield  {author} {\bibinfo {author} {\bibfnamefont {David}\ \bibnamefont
  {Vanderbilt}},\ }\bibfield  {title} {\enquote {\bibinfo {title} {Soft
  self-consistent pseudopotentials in a generalized eigenvalue formalism},}\
  }\href {\doibase 10.1103/PhysRevB.41.7892} {\bibfield  {journal} {\bibinfo
  {journal} {Phys. Rev. B}\ }\textbf {\bibinfo {volume} {41}},\ \bibinfo
  {pages} {7892--7895} (\bibinfo {year} {1990})}\BibitemShut {NoStop}%
\bibitem [{\citenamefont {Corso}(2014)}]{DalCorso2014}%
  \BibitemOpen
  \bibfield  {author} {\bibinfo {author} {\bibfnamefont {Andrea~Dal}\
  \bibnamefont {Corso}},\ }\bibfield  {title} {\enquote {\bibinfo {title}
  {Pseudopotentials periodic table: From h to pu},}\ }\href {\doibase
  10.1016/j.commatsci.2014.07.043} {\bibfield  {journal} {\bibinfo  {journal}
  {Computational Materials Science}\ }\textbf {\bibinfo {volume} {95}},\
  \bibinfo {pages} {337--350} (\bibinfo {year} {2014})}\BibitemShut {NoStop}%
\bibitem [{\citenamefont {Perdew}\ \emph {et~al.}(2008)\citenamefont {Perdew},
  \citenamefont {Ruzsinszky}, \citenamefont {Csonka}, \citenamefont {Vydrov},
  \citenamefont {Scuseria}, \citenamefont {Constantin}, \citenamefont {Zhou},\
  and\ \citenamefont {Burke}}]{PhysRevLett.100.136406}%
  \BibitemOpen
  \bibfield  {author} {\bibinfo {author} {\bibfnamefont {John~P.}\ \bibnamefont
  {Perdew}}, \bibinfo {author} {\bibfnamefont {Adrienn}\ \bibnamefont
  {Ruzsinszky}}, \bibinfo {author} {\bibfnamefont {G\'abor~I.}\ \bibnamefont
  {Csonka}}, \bibinfo {author} {\bibfnamefont {Oleg~A.}\ \bibnamefont
  {Vydrov}}, \bibinfo {author} {\bibfnamefont {Gustavo~E.}\ \bibnamefont
  {Scuseria}}, \bibinfo {author} {\bibfnamefont {Lucian~A.}\ \bibnamefont
  {Constantin}}, \bibinfo {author} {\bibfnamefont {Xiaolan}\ \bibnamefont
  {Zhou}}, \ and\ \bibinfo {author} {\bibfnamefont {Kieron}\ \bibnamefont
  {Burke}},\ }\bibfield  {title} {\enquote {\bibinfo {title} {Restoring the
  density-gradient expansion for exchange in solids and surfaces},}\ }\href
  {\doibase 10.1103/PhysRevLett.100.136406} {\bibfield  {journal} {\bibinfo
  {journal} {Phys. Rev. Lett.}\ }\textbf {\bibinfo {volume} {100}},\ \bibinfo
  {pages} {136406} (\bibinfo {year} {2008})}\BibitemShut {NoStop}%
\end{thebibliography}%


%

	\clearpage
	\setcounter{figure}{0}
	\renewcommand{\thefigure}{S\arabic{figure}}
	\renewcommand{\theequation}{S\arabic{equation}}
	
	\appendix

	\subsection{Zero density lines}
	\begin{figure}
		\centering
		\includegraphics[width=1\textwidth]{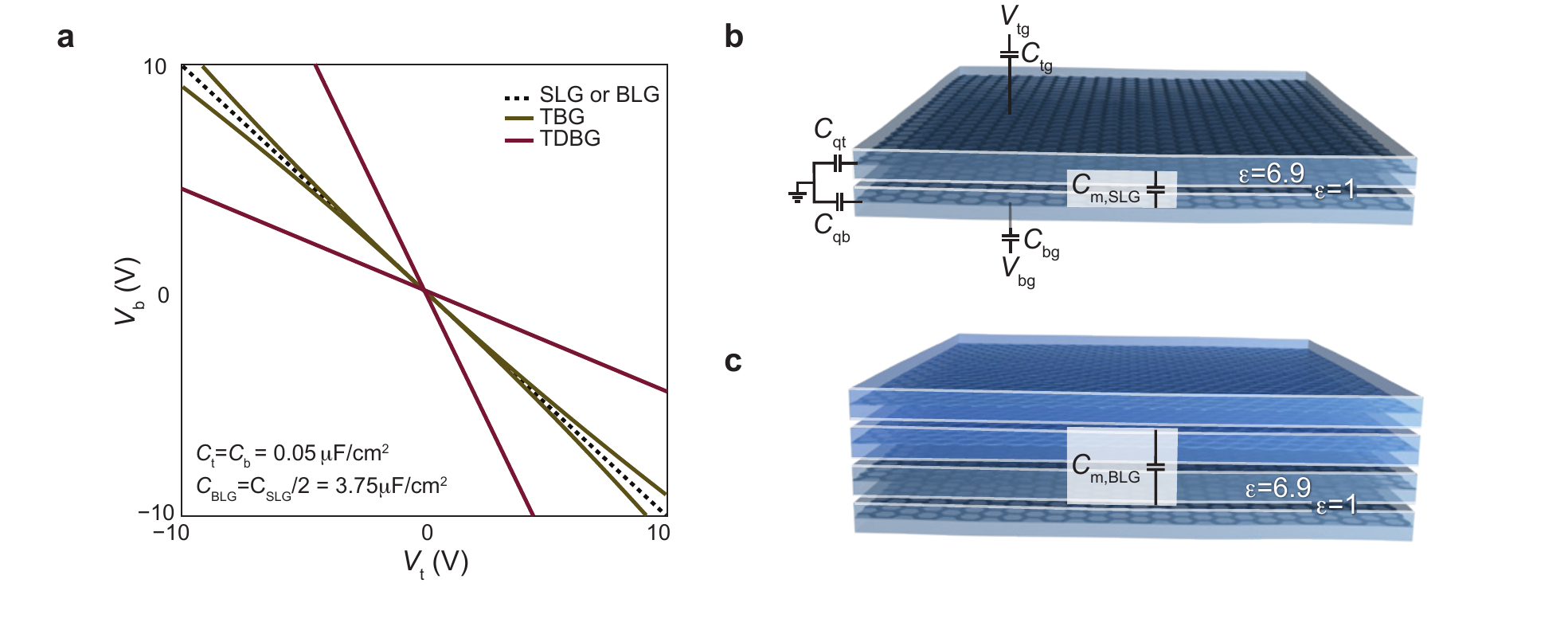}
		\caption{a) Zero-density lines for single-layer graphene (SLG, dashed line), bilayer graphene (BG, dashed line), large-angle twisted bilayer graphene (TBG, brown), large-angle twisted double bilayer graphene (TDBG, red)
			b) Electrostatic model for large-angle twisted bilayer graphene and
			c) large-angle twisted double bilayer graphene, where the geometric capacitance $C _{\rm{m,BG}}=C _{\rm{m,SLG}}/2$.
		}\label{suppfig:splitting}
	\end{figure}
	
	Following the argumentation of Ref.\cite{Rickhaus2019} and using that $\Cb/\Cm\ll1$ and $\Ct/\Cm\ll1$, the zero density lines in the gate-gate map have slopes:
	
	\begin{eqnarray}
	\left.\frac{\partial\VT}{\partial\VB}\right|_{\nb=0}  
	&= -\frac{\Cb}{\Ct}\left(1+\frac{\Ct}{\Cm}+\frac{\Cqt}{\Cm}\right) 
	\approx  -\frac{\Cb}{\Ct}\left(1+\frac{\Cqt}{\Cm}\right) \\
	\left.\frac{\partial\VB}{\partial\VT}\right|_{\nt=0}  
	&= -\frac{\Ct}{\Cb}(1+\frac{\Cb}{\Cm} + \frac{\Cqb}{\Cm})
	\approx  -\frac{\Ct}{\Cb}\left(1 + \frac{\Cqb}{\Cm}\right)\\
	\left.\frac{\partial\VB}{\partial\VT}\right|_{\ntot=0}  
	&= -\frac{\Ct}{\Cb}\cdot\frac{1+\Cb\Cqb + \Cm (\Cqb + \Cqt)}{1+\Ct \Cqt+\Cm (\Cqb + \Cqt) }
	\approx  -\frac{\Ct}{\Cb}
	\end{eqnarray}
	
	In Fig. \ref{suppfig:splitting}a we show the impact of these formulas for different structures in a symmetric geometry. For single- (SLG) and Bernal bilayer graphene (BG), the zero density line consists of a single line with slope $-\Ct/\Cb$. For BG, the applied perpendicular electric field along this line is changing, and it is zero at the origin. A splitting of the zero density line can be observed for large-angle twisted BG, with the amount of splitting given by $(1+\Cq/\Cm)$. The behavior is non-linear, since $\Cq\propto\sqrt{n}$. This is different for large angle TDBG where $\Cq\propto \rm{const.}$ (red lines in the figure). 
	For the depicted lines in the figure we used a symmetric configuration with $\Ct=\Cb=\SI{0.05}{\mu F/cm^2}$. As geometric capacitances between the twisted layers we use $C _{\rm{m,BG}}=C _{\rm{m,SLG}}/2=\SI{3.75}{\mu F/cm^2}$ and for the quantum capacitances standard values from literature. 
	
	In Fig. \ref{suppfig:splitting}b we schematically draw the electrostatic model for large-angle twisted bilayer graphene. The geometric capacitance between the layers, $C _{\rm{m,SLG}}$, can be obtained by considering two "thick" graphene layers with relative dielectric constant $\epsilon_{\rm{g}}=6.9$ and thickness $\dBG=\SI{2.6}{\angstrom}$ separated by vacuum ($\epsilon=1$) and $d=\SI{0.8}{\angstrom}$. In this case:
	\begin{eqnarray*}
		C _{\rm{m,SLG}} = \frac{\epsilon_{\rm{g}}\eps}{\dBG}+\frac{\eps}{d}
	\end{eqnarray*}
	In the case of TDBG, between the center of charges of the top BG and the bottom BG there are two graphene layers and two times a gap with vacuum, therefore
	\begin{eqnarray*}
		C _{\rm{m,BG}} = \frac{\epsilon_{\rm{g}}\eps}{2\dBG}+\frac{\eps}{2d}=\frac{C _{\rm{m,SLG}} }{2}
	\end{eqnarray*}

	\subsection{Extracting the slopes from the measurements}
	\begin{figure}
		\centering
		\includegraphics[width=1\textwidth]{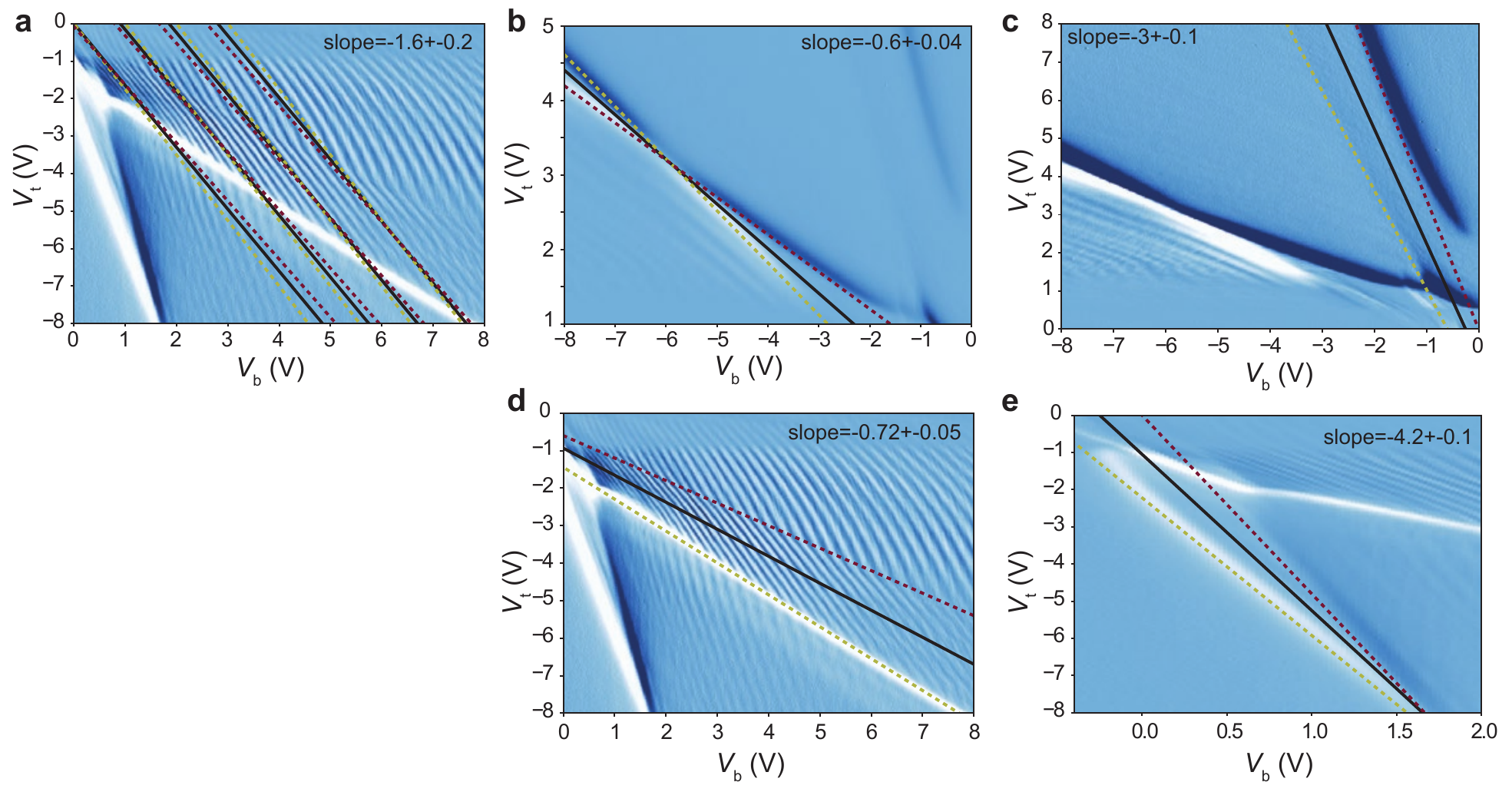}
		\caption{Extracting the slopes of the zero density lines
			\label{suppfig:slopes}	}
	\end{figure}
	Here we describe our procedure to extract the slopes of the zero density condition from the measurement. In Fig.\ref{suppfig:slopes}a we consider the slope of the Fabry-Pérot oscillations of the bottom layer if the top layer is gapped. We find
	\begin{eqnarray*}
		\left.\frac{\partial\VB}{\partial\VT}\right|_{\ntot=0}  
		\approx -\frac{\Ct}{\Cb}  \approx  -1.6\pm0.2\\
	\end{eqnarray*}
	which is in reasonable agreement with the expected slope from electrostatic considerations considering the measured thickness of top and bottom hBN i.e.\ $\Ct/\Cb=\dB/\dT=1.5$. In Figs.\ref{suppfig:slopes}b-e we determine the slopes of the zero density lines by fitting a line in the middle of the gap and we find:
	\begin{eqnarray*}
		\left.\frac{\partial\VB}{\partial\VT}\right|_{\nt=0,\nb<0}   &\approx  -0.6\pm0.04, \;\;\;\;\;
		\left.\frac{\partial\VB}{\partial\VT}\right|_{\nt=0,\nb>0}   &\approx  -0.72\pm0.05 \\
		\left.\frac{\partial\VB}{\partial\VT}\right|_{\nb=0,\nt<0}   &\approx  -4.2\pm0.1, \;\;\;\;\;
		\left.\frac{\partial\VB}{\partial\VT}\right|_{\nb=0,\nt>0}   &\approx  -3\pm0.1 
	\end{eqnarray*}

	\subsection{Electron-hole asymmetry}

	We notice that the slopes of the zero-density lines deviate depending on the charge carrier polarity of the other BG. We attribute this effect to a different effective mass for electrons ($\me$) or holes ($\mh$) which changes the screening. With
	\begin{eqnarray*}
		\Cqte & =&\Cqbe=\Cqe=e^2\Dose=e^2\frac{2\me}{\hbar^2} \;\;\;\;\;\; \mathrm{for\;electrons} \\
		\Cqth & =& \Cqbh=\Cqh=e^2\Dosh=e^2\frac{2\mh}{\hbar^2} \;\;\;\;\;\; \mathrm{for\;holes} 
	\end{eqnarray*}
	we modify the above equations to
	\begin{eqnarray}
	\left.\frac{\partial\VT}{\partial\VB}\right|_{\nb=0,\nt>0}  
	\approx  -\frac{\Cb}{\Ct}\left(1+\frac{\Cqe}{\Cm}\right) \;\;\;\;\;\;
	\left.\frac{\partial\VT}{\partial\VB}\right|_{\nb=0,\nt<0}  
	\approx  -\frac{\Cb}{\Ct}\left(1+\frac{\Cqh}{\Cm}\right) \\
	\left.\frac{\partial\VB}{\partial\VT}\right|_{\nt=0,\nb>0}  
	\approx  -\frac{\Ct}{\Cb}\left(1 + \frac{\Cqe}{\Cm}\right)\;\;\;\;\;\;
	\left.\frac{\partial\VB}{\partial\VT}\right|_{\nt=0,\nb<0}  
	\approx  -\frac{\Ct}{\Cb}\left(1 + \frac{\Cqh}{\Cm}\right)\\
	\label{eq:slopeseh}
	\end{eqnarray}
	
	By using the experimentally extracted value for $\Ct/\Cb$, for the zero-density line of the top BG, we then determine the asymmetry
	\begin{eqnarray*}
		\frac{\Cqe}{\Cqh}=\frac{\me}{\mh}=0.73\pm0.19
	\end{eqnarray*}
	and for the bottom zero density line
	\begin{eqnarray*}
		\frac{\Cqe}{\Cqh}=\frac{\me}{\mh}=0.54\pm0.09
	\end{eqnarray*}
	In average:
	\begin{eqnarray}
	\frac{\me}{\mh}=0.63\pm0.1
	\end{eqnarray}
	Note that this is independent of the value of $\Cm$.

\begin{figure}
		\centering
		\includegraphics[width=1\textwidth]{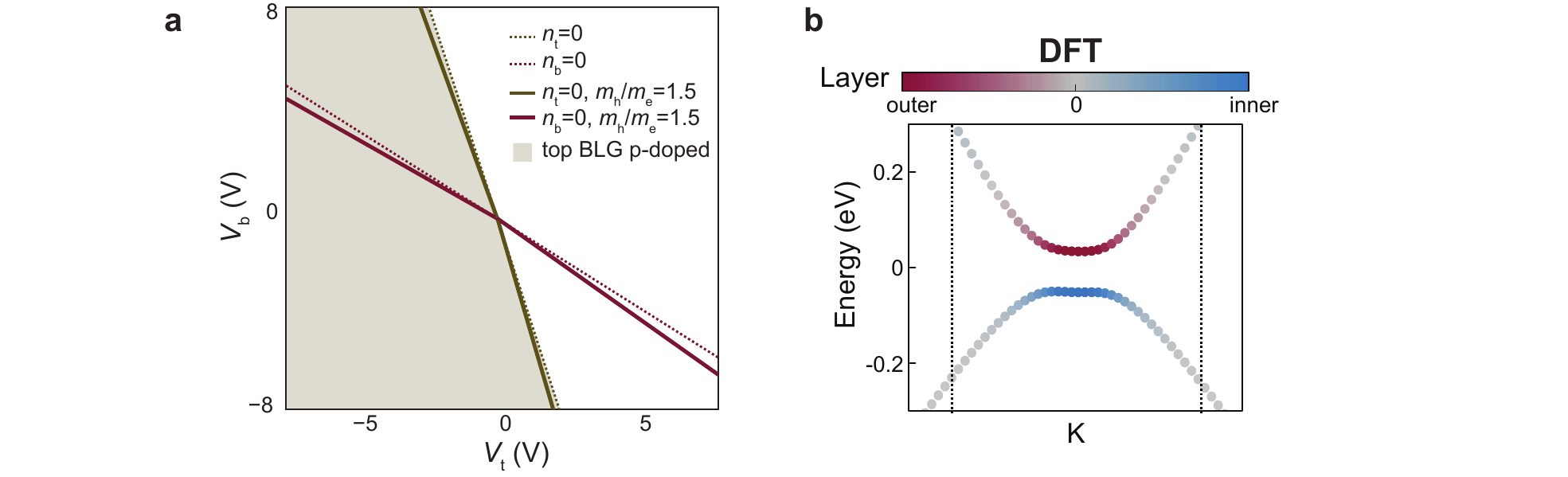}
		\caption{a) Zero density lines if an electron-hole asymmetry is assumed.
			b) Such asymmetry is obviously present in the DFT band structure. Dashed lines are guides to the eye.
		}\label{suppfig:asymmetry}
	\end{figure}

	In Fig.\ref{suppfig:asymmetry}a we show the influence of different electron/hole masses on the zero density lines. In the brown shaded area, the top BG is p-doped. The heavier mass in the valence band leads to a stronger screening of the top-gate. Therefore, a large voltage $\VT$ needs to be applied in order to reach the zero-density line in the bottom layer (red solid line) compared to the case where the masses are equal (red dashed lines). I.e.\, if the top BG is p-doped, the slope of the red solid line is less steep than the red dashed line, and vice-versa if the top BG is n-doped.
	
	In Fig.\ref{suppfig:asymmetry}b we plot again the DFT calculation for low energies (see main text) and added dashed lines as guide to the eyes. Clearly, $\me$ and $\mh$ differ.
	
	\subsection{Estimation of $\Cm$}
	We can estimate the geometric capacitance by making assumptions for the quantum capacitance, i.e.\ $\Cqe=e^2\cdot2\me/\hbar^2\pi$ with $\me=0.80\cdot0.03m_\mathrm{e}$ and $\Cqh=e^2\cdot 2\mh/\hbar^2\pi$ with $\mh=1.26\me$ such that $\me/\mh=0.63$ By inserting this assumption and the measured slopes into Eq.\ref{eq:slopeseh}, we find
	\begin{eqnarray}
	\Cm=3.5\pm\SI{1}{\mu F/cm^2}
	\end{eqnarray} 
	We can estimate whether this number is reasonable by comparing to the measured capacitance between two single-layer graphene sheets, which was $C_\mathrm{m,SLG}=7.5\pm\SI{0.7}{\mu F/cm^2}$ \cite{Rickhaus2019}. Within the error bars, we measure half this capacitance, as expected.

	\section{Determining applied fields to the top/bottom BG}
	
	For just one bilayer system, the displacement field between the two single layers is roughly
	\begin{eqnarray*}
		D=\frac{1}{2\eps}(\Ct\VT-\Cb\VB)
	\end{eqnarray*}
	Where  $\Ct$ ($\Cb$) is the capacitance between the top (bottom) graphene layer and the top (bottom) gate and $\VT$ ($\VB$) the voltage applied to the top (bottom) gate.
	
	For the TDBG system and for zero density in the top two layers (i.e.\ $\nt=0$ and $\Vtop=0$), we notice that the lower two layers act as a gate on the upper layers, i.e.\ we replace $\Cb\VB$ by $\Cm\Vbottom$ (the product of measured interlayer capacitance between the BG sheets $\Cm$ and the electrochemical potential of the lower layer $\Vbottom$) and use that the Fermi energy in the bottom layer is
	\begin{eqnarray*}
		\EFbottom=e\Vbottom=\frac{\hbar^2}{2m^*}\pi\nb=\frac{e^2}{\Cqb}\nb
	\end{eqnarray*}
	Therefore, the displacement field between the upper two layers is:
	\begin{eqnarray}
	D_\mathrm{top}=-\frac{1}{2\eps}(\Ct\VT-\frac{\Cm}{\Cq}\cdot e\nb)
	\end{eqnarray}
	
	We note that we have used two approximations: FIrst, the parabolic dispersion of the bottom BG, i.e.\ no trigonal warping. For large enough $\EFbottom$, this approximation is valid.
	Second, we considered the gating effect of the bottom BG on the top BG and neglected that the bottom BG consists of two layers with different electrochemical potential. Also this approximation is valid for large enough $\EFbottom$.
	
	In analogy we find 
	\begin{eqnarray}
	D_\mathrm{bottom}=\frac{1}{2\epsilon_0}(\Cb\VB-\frac{\Cm}{\Cq}\cdot e\nt)
	\end{eqnarray}
	
	The relation between $D$ and the gap has to be solved self-consistently, but for sufficiently large $D$ it is roughly linear, with a slope of $\SI{195}{meV/(Vnm^{-1})}$.

	\section{Measurements on device 2}
	\begin{figure}
		\centering
		\includegraphics[width=1\textwidth]{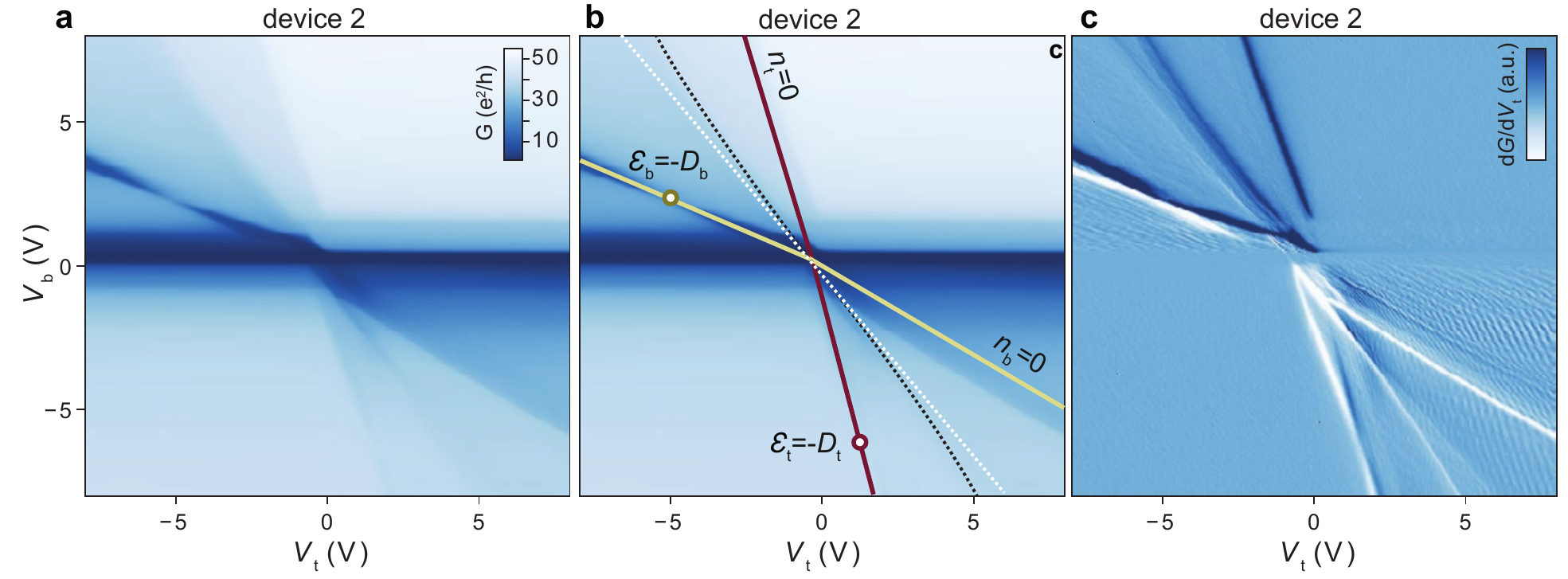}
		\caption{a) $G(\VT,\VB)$ for the second device. 
			b) Similar to the main text, we indicate lines of zero density.
			c) Numerical derivative $dG/d\VT$ of the data.
		}\label{suppfig:device2}
	\end{figure}
	
	We fabricated a second TDBG device which exhibits comparable results. The main differences are:
	\begin{itemize}
		\item The different thickness of top- and bottom hBN ($\dT=\SI{49}{nm}$, $\dB=\SI{56}{nm}$, as compared to device 1 with $\dT=\SI{60}{nm}$, $\dB=\SI{90}{nm}$) results in larger capacitances and thus a stronger gating effect and different slopes in the $G(\VT,\VB)$ maps.
		\item The groundstate resistance is higher, i.e.\ the highest measured resistance in device 2 is $R_\mathrm{max}=\SI{94}{k\Omega}$ compared to $R_\mathrm{max}=\SI{10}{k\Omega}$ in device 1. This may originate from higher quality of the stack.
		\item Unfortunately, the top gate covers an area, where a single-layer flake is twisted on top of a bilayer flake. We call this area twisted single-layer bilayer graphene TSLBG area. The TSLBG area does not connect source and drain contacts, therefore the observation of high resistance is possible. However, the TSBLG area leads to the appearance of a curved zero density line in the conductance map (white dashed line in Fig\ref{suppfig:device2}b, originating from the single-layer graphene flake (compare to Fig.\ref{suppfig:splitting}a) and a straight line (black dashed line in Fig\ref{suppfig:device2}b) that is caused by the bilayer graphene in this area. The presence of the TSBLG are does not affect the analysis of the TDBG features.
	\end{itemize}
	
	We apply the above discussed capacitance model, considering the modified values for $\Ct$ and $\Cb$, and obtain the crystal fields present in this device. We find similar values as for device 1, i.e. 
	\begin{eqnarray*}\cfieldT=\SI{0.11}{V/nm} \\ \cfieldB=\SI{-0.12}{V/nm}\end{eqnarray*}
	With a slope of $\SI{195}{meV/(Vnm^{-1})}$ we find gaps of:
	\begin{eqnarray*}\Delta_{\mathrm{t}}=\SI{21}{meV} \\ \Delta_{\mathrm{b}}=\SI{23}{meV}.\end{eqnarray*}

	\section{Measurements on device 3 with a global topgate}
	\begin{figure}
		\centering
		\includegraphics[width=1\textwidth]{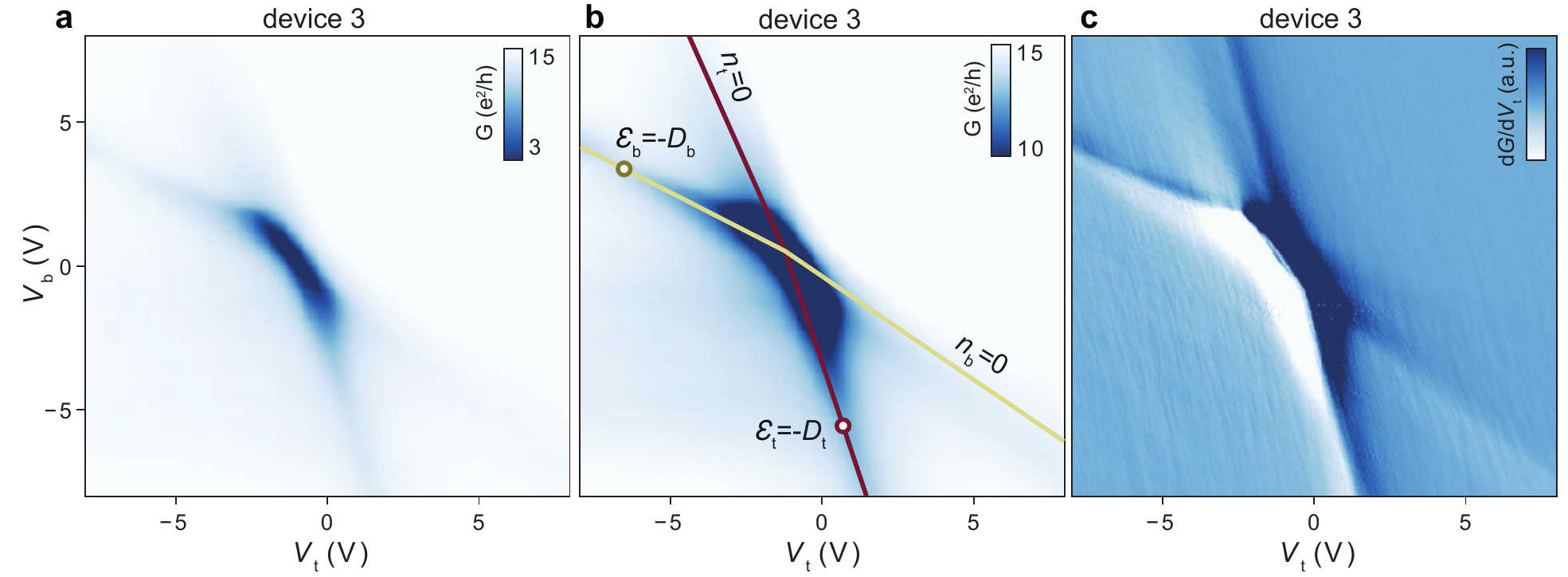}
		\caption{a) $G(\VT,\VB)$ for the third device with a global topgate. 
			b) Device 3 with calculated zero density lines $\nt=0$ and $\nb=0$.
			c) Numerical derivative $dG/d\VT$ of the data.
		}\label{suppfig:device3}
	\end{figure}
	In figure \ref{suppfig:device3} we show the gate-gate map $G(\VT,\VB)$ of a device which has a global topgate, which we realized by evaporating an additional layer of AlO$_x$ after contacting the device and before depositing the top gate. As opposed to devices 1 and 2, the gap appears only in the central region of the plot (around $\VT,\VB=0,0$), i.e. there is no high resistive state present around $\VB=0$ for arbitrary $\VT$, as expected for a global topgate. The zero density lines appear to be more 'blurry' in device 3 compared to devices 2 and 1. We attribute this to the fact that the region that we probe with this device is significantly larger ($2\times\SI{3}{\mu m^2}$) than the region that we probe with the local topgate ($0.4\times{3}{\mu m^2}$), making the conductance more affected by local charge density variations due to disorder. Nevertheless, we do observe charge density lines which are in good agreement with our capacitance model. I.e. the red and yellow line in figure \ref{suppfig:device3}b are obtained by inserting the geometric capacitances of this specific device into equation \ref{eq:slopeseh}. We use $\epsilon_r=9.5$ for the AlO$_x$ layer. At the marked positions of gap closing in the top/bottom layer we find:
	\begin{eqnarray*}\cfieldT=\SI{0.13}{V/nm} \\ \cfieldB=\SI{-0.14}{V/nm}\end{eqnarray*}	With a slope of with a slope of $\SI{195}{meV/(Vnm^{-1})}$ we find gaps of:
	\begin{eqnarray*}\Delta_{\mathrm{t}}=\SI{25}{meV} \\ \Delta_{\mathrm{b}}=\SI{27}{meV}.\end{eqnarray*}

	\section{A model for the Fabry-Pérot resonances}
	\label{FPcalculation}
\begin{figure}
		\centering
		\includegraphics[width=\linewidth]{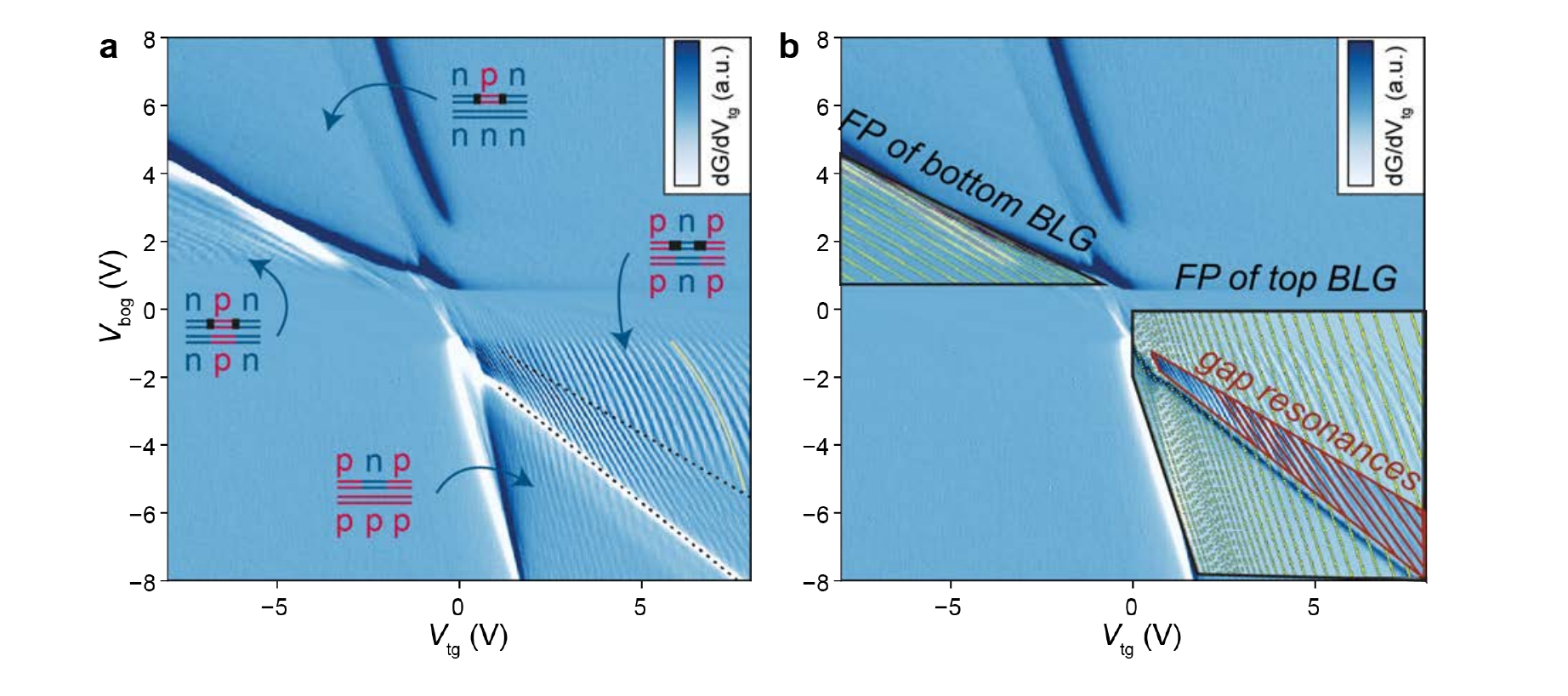}
		\caption{\textbf{a} Numerical derivative of the measured conductance as a function of the gate voltages. Insets show the charge configurations of the device (p = holes, n = electrons). The black dashed lines delimit a region that corresponds to the opening of a gap in the bottom BLG (see main text). The bending of the yellow line is due to a non-constant cavity size (see main text). \textbf{b} Plots of FP resonances calculated with a capacitance model on top of the experimental measurement. The black lines outline the calculations made with the capacitance model of tBBG while the red lines are calculated with the capacitance model of BLG. Lines with very small spacing are not plotted.}
		\label{fig:FPmeasurement}
	\end{figure}

	To obtain the Fabry-Pérot oscillation periodicity, we calculate the densities using :
	\begin{eqnarray*}
		\nt= a_t \VB + b_t \VT\\
		\nb = a_b \VB + b_b \VT 
	\end{eqnarray*}
	where 
	\begin{eqnarray*}
		a_t = \frac{\Cb\Cm\Cq}
		{(\Cq(\Cq+C_{tg}) + \Cb(\Cm+\Cq+\Ct) + \Cm(2\Cq+\Ct))e}\\
		b_t = \frac{\Cq \Ct (\Cb + \Cm + \Cq)}
		{(\Cq(\Cq+\Ct) + \Cb(\Cm+\Cq+\Ct) + \Cm(2\Cq+\Ct))e}\\
		a_b = \frac{\Cq \Cb (\Cm + \Cq + \Ct)}
		{(\Cq(\Cq+\Ct) + \Cb(\Cm+\Cq+\Ct) + \Cm(2\Cq+\Ct))e}\\
		b_b = \frac{\Cq \Cm \Ct}
		{(\Cq(\Cq+\Ct) + \Cb(\Cm+\Cq+\Ct) + \Cm(2\Cq+\Ct))e}.
	\end{eqnarray*}
	From the resulting density maps we calculte the Fabry-Pérot interference pattern using:
	$$\frac{\lambda}{2} N = L,$$
	where $L = 400$ nm is the width of the top gate and N is a natural number. Therefore, knowing that $\lambda = 2\pi/k$, with $k=\sqrt{\pi n_i}$ (i = t, b), we obtain the condition
	\begin{equation}\label{eq:FPn_condition}
	n_i = \frac{\pi N^2}{L^2}.
	\end{equation} 
	Plotting this condition in a gate-gate map yields the FP pattern in the top and bottom bilayer that are depicted on top of the experimental measurement in figure \ref{fig:FPmeasurement}.
	\\
	Figure \ref{fig:FPmeasurement} represents the numerical derivative of the measured conductance with respect to the top gate $\partial G / \partial V_{tg}$, which allows us to observe the FP resonance pattern. On the upper left corner of the map we recognize FP interferences that belong to the pn-junction in the bottom BLG. They match reasonably well with the resonances calculated for the bottom layer in the previous section, which are plotted on top of the corresponding region in figure \ref{fig:FPmeasurement} b. The same is valid for the top bilayer resonances in the bottom right corner, except for a change in slope (yellow line in figure \ref{fig:FPmeasurement} a) due to an increase of the cavity size at low densities (see e.g. Supporting Information of Ref. \cite{Handschin2017}), which is not captured in our model.
	\\
	A third slope of the interference pattern is observed between the black lines of figure \ref{fig:FPmeasurement} a. This pattern matches reasonably well with the plot of FP resonances in dual gated BLG that we show in red in figure \ref{fig:FPmeasurement} b. These plots were obtained by applying the FP condition (eq \ref{eq:FPn_condition}) using the calculated density of only one BLG. From these results we deduce that the top BLG in this area is not affected by the screening of the bottom BLG, which is possible if the Fermi energy in the bottom BLG is in a gap. This demonstrates that the region of reduced conductance around the zero density lines observed indeed correspond to gapped regions. 
	\\
	The mid-gap states in this region may be responsible for the slight difference between the predicted and the measured interference pattern in the gap: in our calculations we approximate the bottom BLG as a charge neutral layer, while instead it may contain non-conducting mid-gap states which can give a small contribution to the screening.
	\\

	\subsection{Thermal activation behavior of device 1}
	\begin{figure}
		\centering
		\includegraphics[width=1\textwidth]{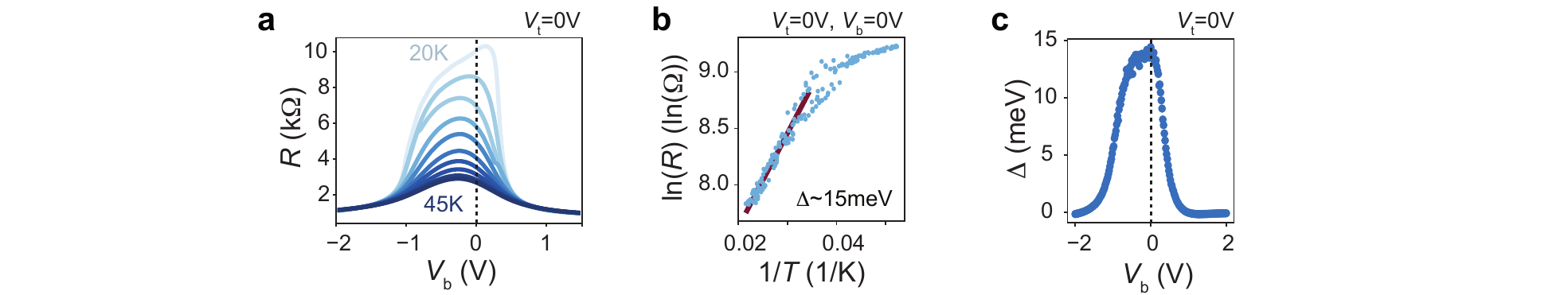}
		\caption{
			a) $R(\VB)$ for $\VT=0$ and temperatures between $\SI{20}{K}$ and $\SI{45}{K}$ reveal the thermal activation of the gap in the ground state of device 1. 
			b) In an Arrhenius plot ($\VT=\VB=0$), the slope at high temperatures (red line) corresponds to a gap of $\Delta=\SI{15}{meV}$. In c) we plot the extracted $\Delta(\VB)$. 
		}\label{suppfig:thermaldev1}
	\end{figure}
	In figure \ref{suppfig:thermaldev1} we show the thermally activated behavior of device 1 (local topgate). We extract the gap as a function of $\VB$ by setting $\VT=0$.

	\section{Measurements on device 3 with a global topgate}
	\begin{figure}
		\centering
		\includegraphics[width=1\textwidth]{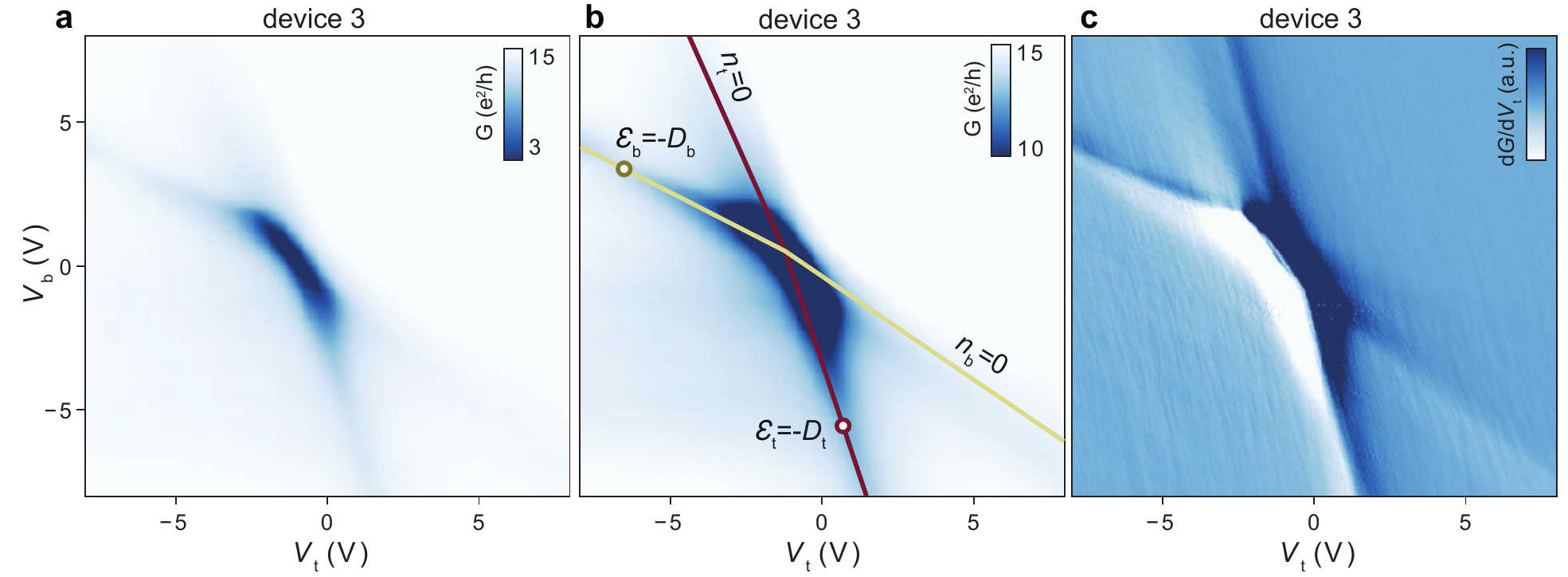}
		\caption{a) $G(\VT,\VB)$ for the third device with a global topgate. 
			b) Device 3 with calculated zero density lines $\nt=0$ and $\nb=0$.
			c) Numerical derivative $dG/d\VT$ of the data.
		}\label{suppfig:device3}
	\end{figure}
	In figure \ref{suppfig:device3} we show the gate-gate map $G(\VT,\VB)$ of a device which has a global topgate, which we realized by evaporating an additional layer of AlO$_x$ after contacting the device and before depositing the top gate. As opposed to devices 1 and 2, the gap appears only in the central region of the plot (around $\VT,\VB=0,0$), i.e. there is no high resistive state present around $\VB=0$ for arbitrary $\VT$, as expected for a global topgate. The zero density lines appear to be more 'blurry' in device 3 compared to devices 2 and 1. We attribute this to the fact that the region that we probe with this device is significantly larger ($2\times\SI{3}{\mu m^2}$) than the region that we probe with the local topgate ($0.4\times{3}{\mu m^2}$), making the conductance more affected by local charge density variations due to disorder. Nevertheless, we do observe charge density lines which are in good agreement with our capacitance model. I.e. the red and yellow line in figure \ref{suppfig:device3}b are obtained by inserting the geometric capacitances of this specific device into equation \ref{eq:slopeseh}. We use $\epsilon_r=9.5$ for the AlO$_x$ layer. At the marked positions of gap closing in the top/bottom layer we find:
	\begin{eqnarray*}\cfieldT=\SI{0.13}{V/nm} \\ \cfieldB=\SI{-0.14}{V/nm}\end{eqnarray*}	With a slope of $\SI{195}{meV/(Vnm^{-1})}$ we find gaps of:
	\begin{eqnarray*}\Delta_{\mathrm{t}}=\SI{25}{meV} \\ \Delta_{\mathrm{b}}=\SI{27}{meV}.\end{eqnarray*}

	\section{Gate maps at different temperatures}
	\begin{figure}
		\centering
		\includegraphics[width=1\textwidth]{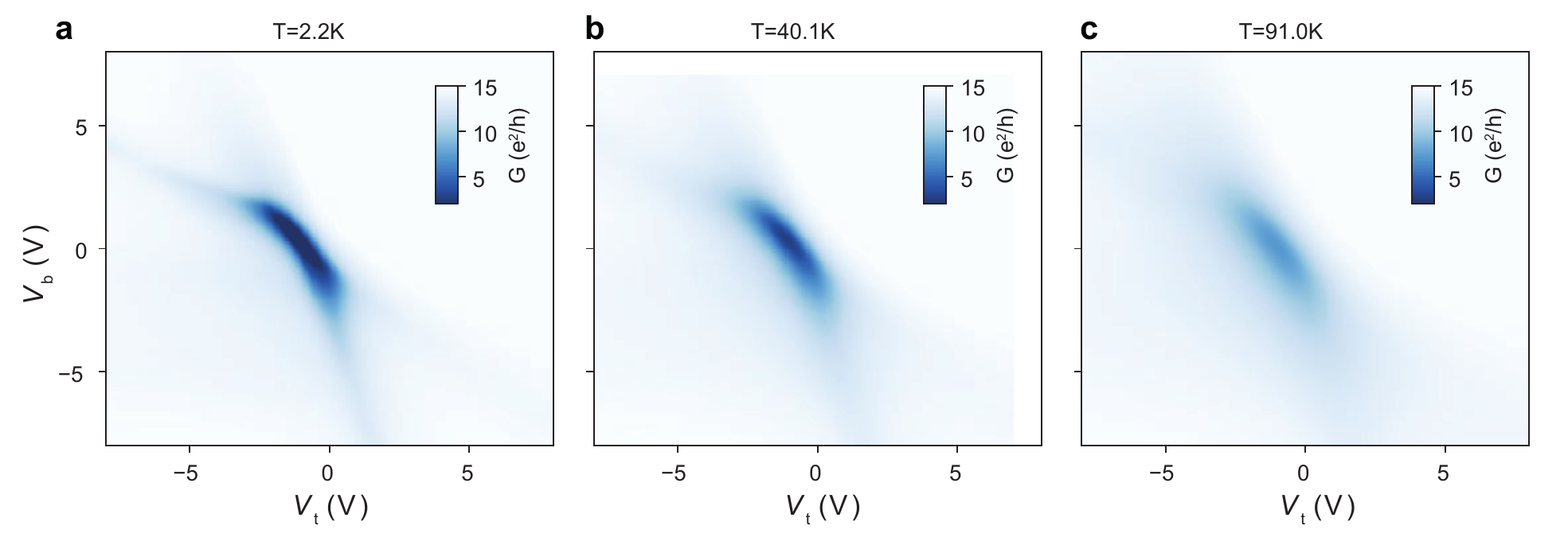}
		\caption{$G(\VT,\VB)$ of device 3 at different temperatures.
		}\label{suppfig:gatemapsT}
	\end{figure}
	In figure \ref{suppfig:gatemapsT} we show the gate-gate map $G(\VT,\VB)$ of device 3 at different temperatures, ranging from $\SI{2.2}{K}$ to $\SI{91}{K}$.

	\section{Density functional theory methods}
	\subsection{Computational details}
	Density functional theory calculations were performed for a structure of twisted double bilayer graphene with a relative rotation angle between the two Bernal stacked bilayers of 13 degrees. The unit cell consists on 152 carbon atoms, and the structure was fully relaxed including van der Waals forces using the Grimme scheme.\cite{Grimme2006,Barone2009} The first principles calculations were performed with the plane-wave pseudopotential formalism as implemented in Quantum Espresso,\cite{Giannozzi2009,Giannozzi2017} using ultrasoft pseudopotentials\cite{PhysRevB.41.7892,DalCorso2014} and PBEsol exchange correlation functional.\cite{PhysRevLett.100.136406}
	
	\begin{figure}
		\centering
		\includegraphics[width=0.7\textwidth]{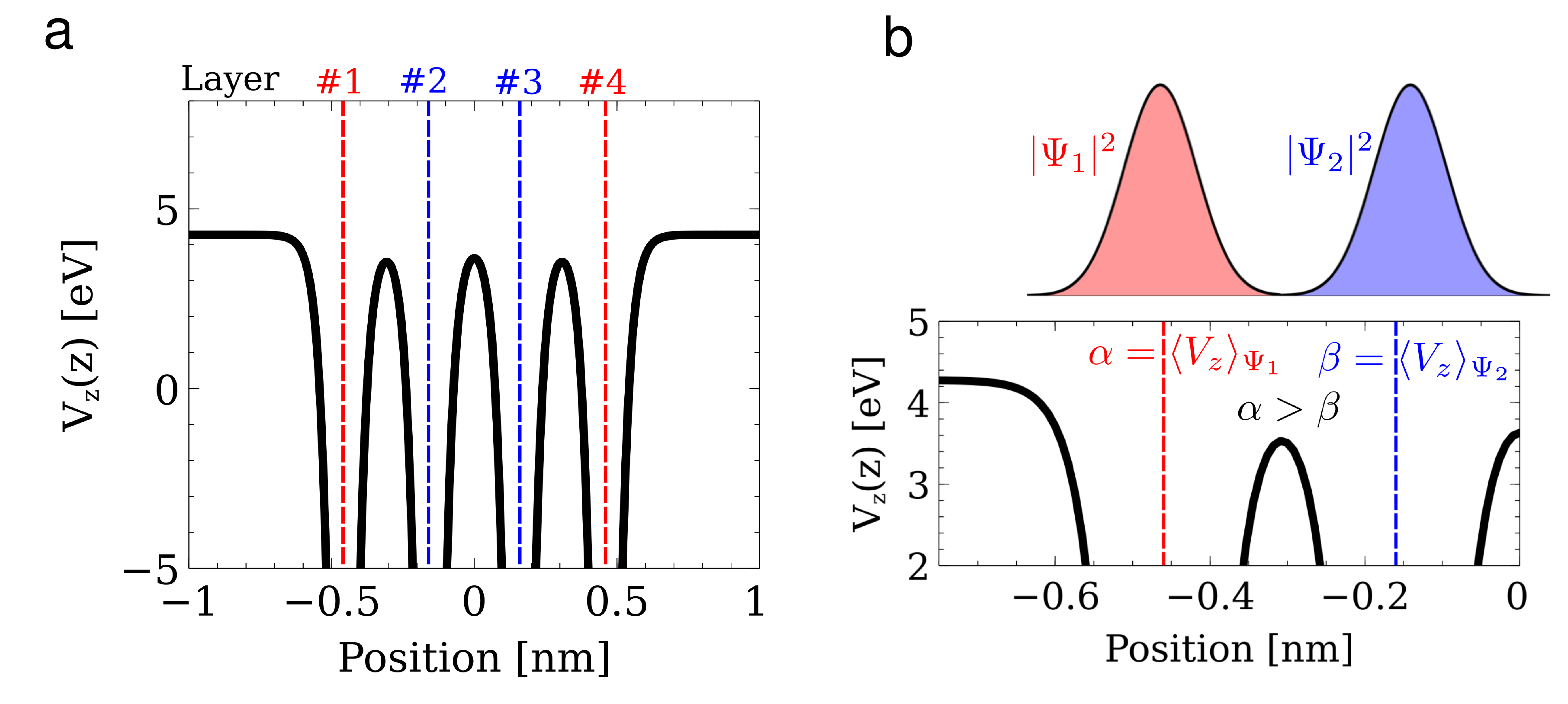}
		\caption{a) Average electrostatic potential in the z-direction as obtained from density functional theory. The dashed lines show the location of each graphene layer, and
			the Fermi energy corresponds to $V_z=0$. Panel
			b) shows a sketch of the origin of the crytal field effect, highlighting that Wannier orbitals in the outer layers will feel a higher electrostatic
			potential, effectively increase the onsite energy in the outer layers and creating a small negative charge in the inner layers.
		}\label{suppfig:dft-potential}
	\end{figure}
	
	\subsection{Microscopic origin of the crystal field effect}
	We now address the microscopic origin of the crystal field contribution in the TDBG using as starting point the first principles results, which can be easily rationalized
	combining the selfconsistent Khon-Sham potential with first order perturbation theory.
	
	We start with $V(x,y,z)$ the electronic potential as obtained from solving the Khon-Sham equations with density functional theory, which gives access to the effective potential
	in each point of the space in the twisted bilayer unit cell. We define the average potential in the $xy$-plane, parallel to the graphene planes as
	
	\begin{equation}
	V_z (z) = \frac{1}{A} \int_{UC} V (x,y,z) dx dy
	\end{equation}
	where $\int_{UC}$ denotes integral of the TDBG unit cell and $A$ is the area of the unit cell in the $xy$ plane. The previous average potential as obtained from the first principles method is shown in Fig. \ref{suppfig:dft-potential}a. We now take a $\Psi_i(z)$ the Wannier wavefunction of an electron localized in layer $i$, where we have integrated out the $xy$ dependence, and we assume to have analogous $z-$profiles for the four different layers. The onsite energies $\alpha,\beta$ in layer 1 and 2 can be computed as
	\begin{equation}
	\begin{matrix}
	\alpha = \int V_z(z) |\Psi_1 (z)|^2 dz \\
	\beta = \int V_z(z) |\Psi_2 (z)|^2 dz
	\end{matrix}
	\end{equation}
	with the onsite difference between the layers $\delta = \alpha - \beta$. It can be easily seen by inspection of the potential profile $V_z$ of Fig. \ref{suppfig:dft-potential}a
	that the integrals $\alpha$
	and $\beta$ will give different results. In particular, since the layer $1$ is closer to the outer region with higher potential, the onsite energy of the outer layers will be higher than
	the one of the inner layers (Fig. \ref{suppfig:dft-potential}b). This yields that for electrons it is energetically favorable to sit in the two inner layers, creating a small internal electric field that yields the inner layers negatively charged and the outer ones positively charged.

	\subsection{Influence of the hBN substrate}
	\subsubsection{Estimate from monolayer graphene encapsulated in hBN}
\begin{figure}
		\centering
		\includegraphics[width=1\linewidth]{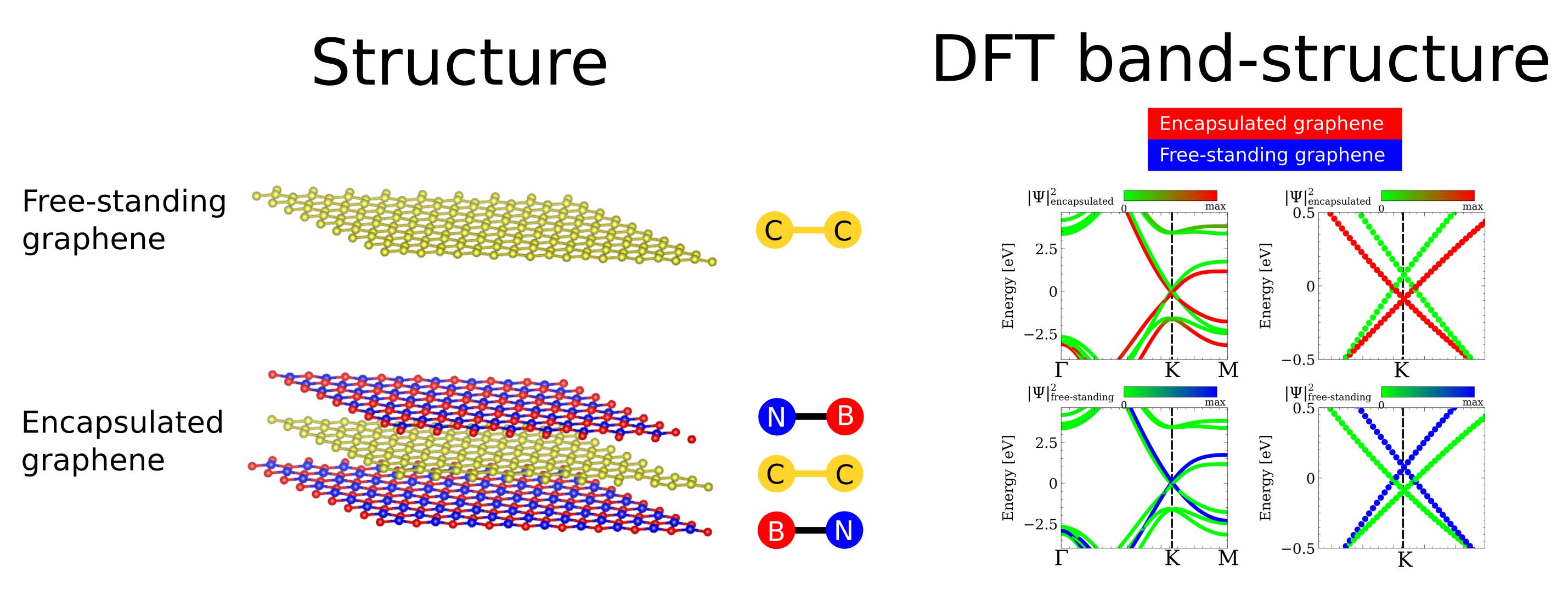}
		\caption{a) Representation of the crystal structure to qualitatively estimate the effect of hBN on the crystal fields. The free-standing graphene is separated by vacuum from the encapsulated graphene and the two systems are decoupled. b) Band structures of the structure represented in figure a), calculated with DFT.}
		\label{fig:hBNeffect}
	\end{figure}

	In the first principle calculations, the top and bottom hBN is neglected in order to reduce the computational cost of the problem. But, because the presence of the hBN lattices influence the crystal field effect, it is relevant to understand their impact. For this reason we compute a structure that allows to estimate their effect on a single layer of graphene, and compare it with the energy dispersion of free standing graphene. Such a structure is schematically represented in figure \ref{fig:hBNeffect}, where the two SLGs are spatially separated by a large distance. In this way we compute a unit cell with two decoupled graphene layers, an encapsulated and a free-standing one, such that we obtain the energy differenece of the two situations (figure \ref{fig:hBNeffect}). The result shows a splitting of $\sim 120$ meV between the Dirac points of the two SLGs, where the encapsulated graphene has a lower energy. From this observation we deduce that one hBN layer can shift the potential of a graphene layer in its vicinity by $\sim 60$ meV. Therefore, 
	each hBN layer reduces the electrostatic potential of a neighboring graphene layer by 60 meV\\ By simply subtracting the previous 60 meV from the gap computed with free-standing tBBG, we estimate $\Delta_0 \approx 20$ meV for encapsulated tBBG, which is of the same order as the experimentally measured gap. Notice that, in order to have a small enough unit cell, we had to stretch the bonds of the hBN lattice.
	\\
	\\
	\subsubsection{Full calculation of twisted double bilayer graphene encapsulated in hBN}
\begin{figure}
		\centering
		\includegraphics[width=0.5\linewidth]{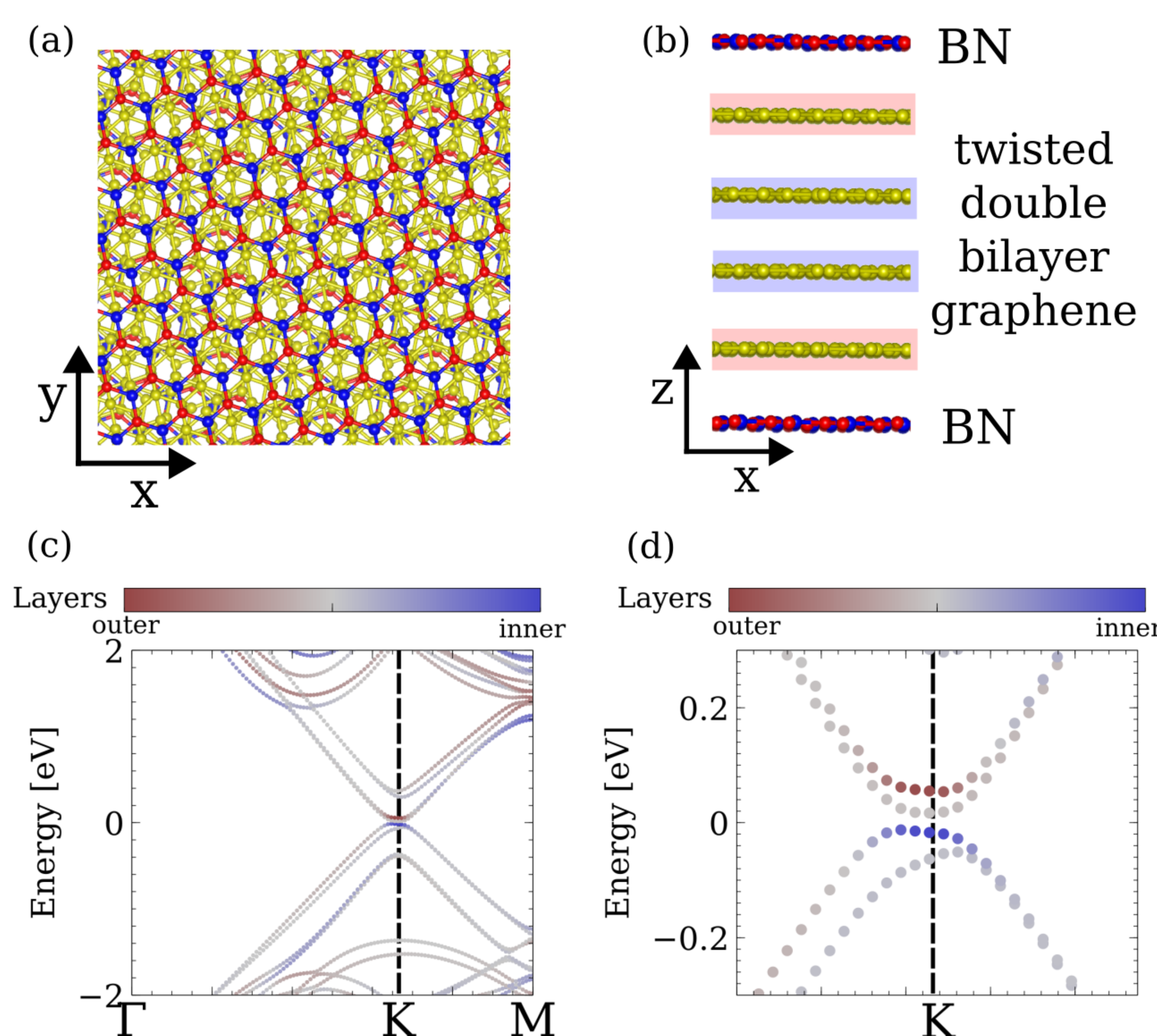}
		\caption{ a,b) Sketch of the twisted double bilayer structure with the hBN encapsulation
			and a rotation angle of 21.8$^\circ$. The first principles band structure c,d) shows an analogous behavior
			to the case without hBN, but presenting a smaller crystal-field induced gap.}
		\label{fig:22hBN}
	\end{figure}

	We now address from first principles a twisted double bilayer together with the hBN encapsulation. Since the calculations involving hBN are computationally expensive, here
	we will focus on an encapsulated double bilayer structure with a rotation angle of 21.8$^\circ$, i.e. a large angle double bilayer which shows an analogous crystal field induced gap to the
	experimental situation. 
	For the sake of completeness, we will compare the electronic structure of such twisted double bilayer with and without the hBN encapsulation, showing that no qualitative change
	is introduced by the hBN substrate.
	\\
	\begin{figure}
		\centering
		\includegraphics[width=0.5\linewidth]{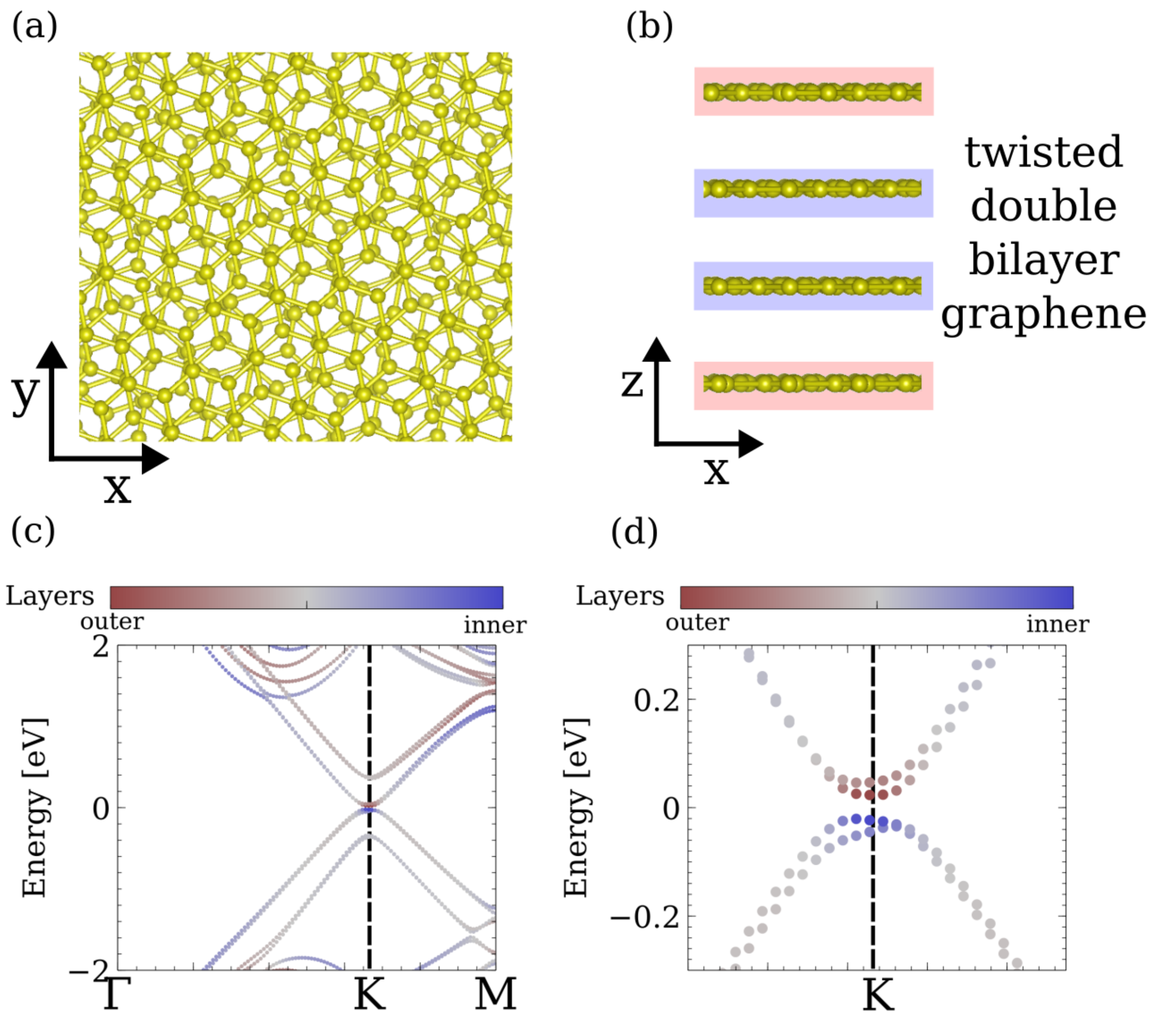}
		\caption{ a,b) Sketch of the twisted double bilayer structure without the hBN encapsulation
			and a rotation angle of 21.8$^\circ$. Panels
			c,d) show the first principles band structure, showing that the top
			of the valence band is localized in the two inner layers, which is associated
			with the crystal field induced gap.}
		\label{fig:22nohBN}
	\end{figure}    
	\\
	We first focus on the twisted double bilayer without hBN encapsulation and a rotation angle of 21.8$^\circ$ (Fig. \ref{fig:22nohBN}a,b). As shown in the first principles
	band structures of Fig. \ref{fig:22nohBN}c,d, an analogous crystal field induced gap appears in the system, creating a small charge imbalance between the inner and outer layers.
	This is the same phenomenology as it was observed in the twisted structure at 13 degrees rotation, which highlights that the crystal field induced gap happens for generic large angles.
	\\
	We now move on to the twisted double bilayer with hBN encapsulation and a rotation angle of 21.8$^\circ$ (Fig. \ref{fig:22hBN}a,b). In this situation, we also observe a crystal field
	induced gap in the encapsulated twisted double bilayer (Fig. \ref{fig:22hBN}c,d), similar to the one found without the encapsulation. The top of the valence band is also localized
	in the inner layers, showing that this encapsulated system will also present a small charge imbalance. The effect of the hBN is to slightly decrease the value of the crystal field
	induced gap, yet without creating a qualitative change in the electronic structure.
	\\

\end{document}